\documentclass[12pt]{article}
\pdfoutput=1
\usepackage{graphicx}
\usepackage{epstopdf}
\usepackage{amsmath}
\usepackage{amsfonts}
\usepackage{amssymb}
\usepackage{color}
\usepackage{mathrsfs}   

\usepackage[font={footnotesize}]{caption}

\pdfoutput=1

\usepackage[english]{babel}

\usepackage[colorlinks,bookmarks]{hyperref}
\definecolor{linkblue}{rgb}{0,0,0.8}
\definecolor{linkgreen}{rgb}{0,0.5,0}

\hypersetup{pdfpagemode=UseNone, pdfstartview=FitH, linkcolor=linkblue, %
            citecolor=linkgreen, urlcolor=linkblue}

\numberwithin{equation}{section}

\definecolor{darkgreen}{rgb}{0,0.3,0}
\definecolor{darkblue}{rgb}{0,0,0.3}
\definecolor{darkred}{rgb}{0.7,0,0}

\newcommand{\be}{\begin{equation}}
\newcommand{\bse}{\begin{subequations}}
\newcommand{\ese}{\end{subequations}}
\newcommand{\bea}{\begin{eqnarray}}
\newcommand{\eea}{\end{eqnarray}}
\newcommand{\ba}{\begin{array}}
\newcommand{\ea}{\end{array}}
\newcommand{\ee}{\end{equation}}

\newcommand{\half}{\frac{1}{2}}

\newcommand{\eqn}[1]{Eq.~(\ref{#1})}

\newcommand{\kvec}{\vec{k}}

\newcommand{\qvec}{\vec{q}}
\newcommand{\momspmeas}[1]{\frac{d^3 #1}{(2 \pi)^3}}

\newcommand{\knl}{k_{\rm NL}}

\newcommand{\unitsk}{\, h { \rm Mpc^{-1}}}  

\newcommand{\Poneloop}{P_{1-\text{loop}}}

\newcommand{\website}{\url{http://web.stanford.edu/~senatore/}}

\newcommand{\ta}{\tilde{a}}

\newcommand{\mG}{\mathcal{G}}

\newcommand{\mU}{\mathcal{U}}
\newcommand{\mV}{\mathcal{V}}

\newcommand{\vk}{\vec{k}}
\newcommand{\vq}{\vec{q}}

\makeatletter \@addtoreset{equation}{section}

\makeatletter\renewcommand\section{\@startsection {section}{1}{\z@}%
                                   {-3.5ex \@plus -1ex \@minus -.2ex}
                                   {2.3ex \@plus.2ex}%
                                   {\normalfont\large\bfseries}}
\renewcommand\subsection{\@startsection{subsection}{2}{\z@}%
                                     {-3.25ex\@plus -1ex \@minus -.2ex}%
                                     {1.5ex \@plus .2ex}%
                                     {\normalfont\bfseries}}

\begin{document}

\begin{flushright}
\end{flushright}

\begin{center}

{\LARGE{\bf IR-safe and UV-safe integrands in the \\ EFTofLSS with exact time dependence}}
\\[0.7cm]
{\large Matthew Lewandowski${}^{1}$,  Leonardo Senatore${}^{2,3}$}
\\[0.7cm]
\vspace{.3cm}
{\normalsize { \sl $^{1}$ Institut de Physique Th\'eorique, Universit\'e Paris Saclay,\\ CEA, CNRS, 91191 Gif-sur-Yvette, France}}\\

\vspace{.3cm}
{\normalsize { \sl $^{2}$ Stanford Institute for Theoretical Physics,\\ Stanford University, Stanford, CA 94306}}\\
\vspace{.3cm}

{\normalsize { \sl $^{3}$ Kavli Institute for Particle Astrophysics and Cosmology, \\
Physics Department and SLAC, Menlo Park, CA 94025}  }\\

\vspace{.3cm}


\end{center}

\vspace{8mm}
\setcounter{page}{1} \baselineskip=15.5pt \thispagestyle{empty}

\hrule \vspace{0.3cm}

{\small  \noindent \textbf{Abstract} 

\vspace{.1in}

\noindent  Because large-scale structure surveys may very well be the next leading sources of cosmological information, it is important to have a precise understanding of the cosmological observables; for this reason, the Effective Field Theory of Large-Scale Structure (EFTofLSS) was developed.  So far, most results in the EFTofLSS have used the so-called Einstein-de Sitter approximation, an approximation of the time dependence which is known to be accurate to better than one percent.  However, in order to reach even higher accuracy, the full time dependence must be used.  The computation with exact time dependence is sensitive to both infrared (IR) and ultraviolet (UV) effects in the loop integrands, and while these effects must cancel because of diffeomorphism invariance, they make numerical computation much less efficient.  We provide a formulation of the one-loop, equal-time exact-time-dependence power spectrum of density perturbations which is manifestly free of these spurious IR and UV divergences at the level of the integrand.  We extend our results to the total matter mode with clustering quintessence, show that IR and UV divergences cancel, and provide the associated IR- and UV-safe integrand.  This also establishes that the consistency conditions are satisfied in this system.  We then use our one-loop result to do an improved precision comparison of the two-loop dark-matter power spectrum with the Dark Sky $N$-body simulation.
}

 \vspace{0.5cm}
\hrule

 \vspace{0.3cm}

\newpage

\tableofcontents


%
%

\section{Introduction} \label{introsec}

Large-scale structure surveys may potentially be our next leading source of cosmological information because the amount of information in such surveys scales roughly like $k_{\rm max}^3$, where $k_{\rm max}$ is the largest wavenumber under theoretical control.  Thus, it is important to have a precise understanding of large-scale structure (LSS) observables at the highest $k$ (most ultraviolet, or UV) possible.  In order to address this concern, the Effective Field Theory of Large-Scale Structure (EFTofLSS) was developed (there is by now a rather large literature, see for example~\cite{Baumann:2010tm,Carrasco:2012cv,Porto:2013qua,Senatore:2014via,Carrasco:2013sva,Carrasco:2013mua,Pajer:2013jj,Carroll:2013oxa,Mercolli:2013bsa,Angulo:2014tfa,Baldauf:2014qfa,Senatore:2014eva,Senatore:2014vja,Lewandowski:2014rca,Mirbabayi:2014zca,Foreman:2015uva,Angulo:2015eqa,McQuinn:2015tva,Assassi:2015jqa,Baldauf:2015tla,Baldauf:2015xfa,Foreman:2015lca,Baldauf:2015aha,Baldauf:2015zga,Bertolini:2015fya,Bertolini:2016bmt,Assassi:2015fma,Lewandowski:2015ziq,Cataneo:2016suz,Bertolini:2016hxg,Fujita:2016dne,Perko:2016puo,Lewandowski:2016yce}) to describe gravitational clustering in the mildly non-linear regime through the correct treatment of the effects of UV modes on large-scale observables.  The main idea is that, in order to correct mistakes introduced in perturbation theory from uncontrolled short-scale physics, one must include suitable counterterms in the perturbative expansion.  Once the coefficients, or coupling constants, of these counterterms are fit to observation, they will correctly describe the effects of short-scale physics on the large-scale modes that we directly observe in large-scale structure surveys.  The result is a controlled expansion in $k / \knl$, where $\knl$ is the scale at which the effective field theory can no longer describe the dynamics (i.e. it is the strong coupling scale).  For $k \lesssim \knl$, predictions can be computed to arbitrary precision (up to non-perturbative effects) by including more and more loops.  So far, this research program has shown that large-scale structure clustering can be accurately described for dark matter \cite{Foreman:2015lca}, galaxies \cite{Senatore:2014eva, Angulo:2015eqa}, including baryons \cite{Lewandowski:2014rca}, and in redshift space \cite{Perko:2016puo, Lewandowski:2015ziq} up to relatively high wavenumbers.  Codes used to produce some of the results mentioned in the former references, as well as the one used to obtain the results presented in this paper, are publicly available online.\footnote{\website}

When constructing the perturbative solution, loop integrals need to be performed.  For IR-safe (IR stands for infrared) quantities, the loop integrals have spurious IR divergencies that cancel in the final answer (this cancellation being guaranteed by diffeomorphism invariance~\cite{Carrasco:2013sva}).  As it was noted in~\cite{Carrasco:2013sva}, the numerical cost of the computation (as well as the conceptual cleanliness of it) can be ameliorated by constructing an IR-safe integrand.  Reference \cite{Carrasco:2013sva} constructed the IR-safe integrand in the Einstein-de Sitter (EdS) approximation, where all loop contributions have the same time dependence (for example, all $1$-loop contributions are proportional to $D_+(a)^4$, where $D_+(a)$ is the linear growth factor).  This treatment made manifest the cancellation of large IR contributions in the loop expansion, which are guaranteed to cancel by the equivalence principle in equal-time correlation functions of density perturbations, at the level of loop \emph{integrands}; practically, this means that the computer does not waste time precisely computing very large numbers which will ultimately cancel each other to give a much smaller result.  These cancellations were originally studied in~\cite{Jain:1995kx, Scoccimarro:1995if}, and more recently in~\cite{Peloso:2013zw}.  The IR properties of the loop expansion are related to the so-called consistency conditions for dark matter \cite{Jain:1995kx, Peloso:2013zw, Scoccimarro:1995if, Kehagias:2013yd, Bernardeau:2001qr, Bernardeau:2011vy, Bernardeau:2012aq}, which~\cite{Carrasco:2013sva, Creminelli:2013mca} pointed out are also a result of the equivalence principle (\cite{Creminelli:2013mca} used an explicit construction of adiabatic modes due to Weinberg \cite{Weinberg:2003sw}).  The fact that there was only one mode (dark matter) present in these discussions was important because it meant that there was a unique freely falling observer to transform to and remove not only gradients of the gravitational potential (which is always possible), but also the velocity of the species.  The fact that large-scale velocity effects do not cancel in equal-time correlations functions in the presence of multiple modes was pointed out in~\cite{Tseliakhovich:2010bj, Bernardeau:2011vy, Bernardeau:2012aq}.

A similar scenario exists for spurious UV contributions to loop integrals. The leading behavior in the UV ($k/q \rightarrow 0$) is fixed to be
\begin{align} \label{p13uvv}
P_{13}^{\rm UV} ( k ) & \propto  k^2 P_{11}(k) \int \momspmeas{q}  \frac{1}{q^2} P_{11}(q) \ ,  \\ 
P_{22}^{\rm UV} ( k ) & \propto k^4 \int \momspmeas{q} \frac{1}{q^4} P_{11} ( q ) P_{11} ( q ) \ , \label{p22uvv}
\end{align}
which are indeed the kinds of terms which can be corrected by counterterms in the perturbative expansion.  When computing with exact time dependence, though, $P_{13}$ and $P_{22}$ are each a sum of different terms with different time dependences.  These terms happen to have individually spurious leading UV behavior, which must cancel separately in $P_{13}$ and $P_{22}$ (because they have different $k$ dependences).  For exactly the same reasons mentioned above for the spurious IR terms, we would like to eliminate the spurious UV terms at the level of the integrand.  This is exactly what we do in this paper: we provide a manifestly IR- and UV- safe (which we will abbreviate as IR\&UV-safe for now on) integrand.

In this paper, we extend the results of~\cite{Carrasco:2013sva} to provide the one-loop IR\&UV-safe integrand with exact time dependence, i.e. when the time dependence is computed with the exact Green's functions.  To date, most of the computations in the EFTofLSS have been done in the EdS approximation, and because we ultimately want the most precise computation possible, it may soon become necessary to use the exact time dependence.  It was shown \cite{Takahashi:2008yk} (see also \cite{martel1991, Bernardeau:1993qu, Scoccimarro:1997st, Carrasco:2012cv} for related discussions) that the difference between the EdS approximation and exact time dependence in $\Lambda$CDM at $z=0$ for the total power spectrum up to one loop is about $0.5 \%$.  A potential problem with using the exact time dependence, though, is that the more complex expressions in terms of Green's functions obscure the cancellation of IR and UV divergences, which nevertheless must happen because of diffeomorphism invariance.  In this paper, we explicitly show how these cancellations come about and then construct a one-loop integrand which is manifestly free of these large spurious IR and UV terms.  While the two different loop contributions, $P_{22}( a, k)$ and $P_{13}(a,k)$, have different time dependences, we show that the leading IR and UV terms in fact do have the same time dependence, thus allowing the spurious IR and UV terms to cancel.  Because of the delicate cancellation of terms which generally have different time dependences, a numerical computation of the loops with exact time dependence is more sensitive to the precision with which the time-dependent factors are computed; a small error in the relative value of the time-dependent coefficients means that numerically the spurious IR and UV terms will not exactly cancel, and because they are proportional to large factors (this is indeed what it means to be IR or UV divergent) the overall numerical error can be high.\footnote{In practice, because the real universe has a natural IR cutoff around the matter-radiation equality scale, $k_{\text{eq}} \sim 0.01 \unitsk$, these IR terms are not too large in the one-loop computation that we present here. In fact, as we will see below, it is more important to remove spurious UV terms.  However, the IR terms are expected to be much more of a nuisance in a two-loop or higher order computation  than in our one-loop study.}   Said another way, the IR and UV divergent terms are only guaranteed to cancel if the numerical coefficients are computed with infinite precision.  This situation could be counterproductive.  The reason that we are interested in computing with exact time dependence is because we want to compute observables with the most precision possible, so the precision lost due to IR and UV effects better not outweigh the precision gained in using the exact time dependence.  In this paper, we provide an expression for the one-loop power spectrum with exact time dependence which is immune to these concerns by writing it in terms of a manifestly IR\&UV-safe integrand, where there are no spurious IR or UV divergences to be cancelled numerically.  Then, we use our results to do an improved precision comparison of the two-loop dark-matter power spectrum (now with exact time dependence, and IR\&UV-safe integrand used for $P_{1-\text{loop}}$) with the Dark Sky $N$-body simulation~\cite{Skillman:2014qca}.

It is interesting to note (as we will show later) that our results here can be easily extended to another system of interest: dark matter coupled to quintessence with zero speed of sound (also called clustering quintessence), which was a main example used in~\cite{Lewandowski:2016yce} to describe how to include dark energy (in the so-called Effective Field Theory of Dark Energy formalism) in the EFTofLSS, and had been previously studied outside of the context of the EFTofLSS in~\cite{Creminelli:2008wc, Creminelli:2009mu, Sefusatti:2011cm}.  There, the equations of motion for the adiabatic mode (called $\delta_A$) are the same as those in $\Lambda$CDM apart from a different time dependent factor in the continuity equation (see \eqn{lineareqs} and \eqn{lineareqs2}).  In this paper we also show that for equal-time correlation functions of the adiabatic mode in the dark matter and clustering quintessence system, the individual IR and UV divergent terms of the one-loop contribution cancel, and we provide the IR\&UV-safe integrand for that system.  Indeed, it is not surprising that this is the case.  When the quintessence has a small speed of sound, i.e. $c_s^2 \rightarrow 0$, the system reduces to a single mode $\delta_A$ (the isocurvature mode is proportional to $c_s^2$, and thus is absent when $c_s^2 \rightarrow 0$), and because the equivalence principle is not violated, the arguments of~\cite{Carrasco:2013sva} apply.  In particular, the fact that the spurious IR terms cancel also means that the consistency conditions are still satisfied for correlation functions of $\delta_A$.  However, when $c_s^2 \neq 0$ but $| c_s^2| \ll 1$, an isocurvature mode proportional to $c_s^2$ is generated \cite{Lewandowski:2016yce}, and we expect the loop integrals to have a strong IR dependence proportional to $c_s^2$, and the consistency conditions to be violated by terms proportional to $c_s^2$.

This paper is organized as follows.  In the beginning of Section~\ref{irsafecc} we review the construction of the IR-safe integrand with approximate time dependence presented in~\cite{Carrasco:2013sva}.  In Section~\ref{oneloopkernelssec}, we present the loop contributions with exact time dependence, $P_{13} ( a , k)$ and $P_{22} ( a , k)$, in a way to prepare us for the computation, and in Section~\ref{irlimitsec} we find the IR limit of our expressions.  Then, in Section~\ref{irsafeintegrandsec} we first show how the leading IR terms cancel, and then we construct the IR-safe integrand, which gives an expression for the integrand which is manifestly free of IR divergences at every step of the numerical computation.  In Section~\ref{uvsafesec}, we do the same, but for the spurious UV terms.  In Section~\ref{resultssec} we present some numerical results of our study, and in Section~\ref{preccompsec} we present our results of the improved precision comparison.  Finally, in Section~\ref{conclusionsec}, we conclude.

%
%
\section{Review of IR-safe integrand with EdS approximation}   \label{irsafecc}

Let us first look at the case previously studied in \cite{Carrasco:2013sva} (see also~\cite{Blas:2013bpa}), which uses the EdS approximation for the loop integrals.  In that case, we have 
\be
P_{1-\text{loop}} (a,k)  = P_{22} ( a,k ) + P_{13}(a,k) \ ,
\ee
where the respective integrands are defined as 
\be
P_{22} ( a,k) = \frac{D(a)^4}{D(a_i)^4} \int \momspmeas{q} \, p_{22} ( \kvec , \qvec)  \hspace{.3in} \text{and} \hspace{.3in} P_{13} (a, k ) =  \frac{D(a)^4}{D(a_i)^4}  \int \momspmeas{q} \, p_{13} ( \kvec , \qvec) \ ,  
\ee
so that for equal time power spectra, we only need to consider the momentum dependence in the functions $p_{22} ( \kvec , \qvec)$ and $p_{13} ( \kvec , \qvec)$ to examine the IR properties.  While the explicit forms of $p_{22} ( \kvec , \qvec)$ and $p_{13} ( \kvec , \qvec)$ can be found in \cite{Carrasco:2013sva}, we will only need the forms in the IR limits $q / k \rightarrow 0$ and $\qvec \rightarrow \kvec$.  In the $q / k \rightarrow 0$ limit, and taking the initial power spectrum to be $P_{11} ( k ) \propto k^n$, we have 
\begin{align} \label{p13div}
p_{13} ( \kvec , \qvec) & \underset{q/k \rightarrow 0}{\sim} - k^{n + 2} \mu^2 q^{n-2} + \mathcal{O}( q^{n} ) \ ,  \\
 p_{22} ( \kvec , \qvec) &  \underset{q/k \rightarrow 0}{\sim} \half k^{n + 2} \mu^2 q^{n-2} + \mathcal{O}( q^{n-1} ) \label{p22div}\ ,
\end{align}
where $\mu \equiv \kvec \cdot \qvec / ( k \, q)$.  The equivalence principle ensures that the effect of the loop from the IR must start like $\int d^3 q \, P_{11} ( q )$, but looking at \eqn{p13div} and \eqn{p22div}, there seem to be terms which go like $\int d^3 q \, (k/q)^{2} P_{11}(q)$ and $\int d^3 q \, (k/q) P_{11} ( q)$ (which we will generically refer to as divergent terms), so the cancellation is not manifest.  Being more careful though, one notices that there is another IR-divergence in $p_{22}$ which comes from the limit $\qvec \rightarrow \kvec$ (i.e. sending the other leg of the loop momentum to zero), which goes like 
\be
  p_{22}( \kvec , \qvec) \underset{\qvec \rightarrow \kvec}{\sim} \half k^n \frac{ ( \kvec \cdot [ \kvec - \qvec])^2}{ q^2 |\kvec - \qvec|^2}  |\kvec - \qvec|^n \ . 
\ee
Summing the above IR divergences, it was found that the leading divergence which goes like $ d^3 q \, (k/q)^{2} P_{11}(q)$ cancels (as it should), but that the divergences come from two different regions within the integration limits.  Additionally, the divergence in \eqn{p22div} proportional to $ d^3 q \, (k/q) P_{11} ( q)$ is also proportional to $\mu$, and so indeed cancels in the final integration over $d^3 q$.  Thus, if one were to simply add $p_{22} ( \kvec , \qvec) + p_{13} ( \kvec , \qvec)$ and then integrate over $d^3 q$ numerically, the numerical integration would be computing very large numbers near $\qvec \rightarrow 0$ and $\qvec \rightarrow \kvec$ which when summed give a result which is much smaller than the individual numbers computed.  This was known to happen~\cite{Jain:1995kx, Scoccimarro:1995if}, although it was sometimes incorrectly attributed to the Galilean invariance of the Newtonian equations (it is actually guaranteed by diffeomorphism invariance).  All in all, \cite{Carrasco:2013sva} found that the way to write the one-loop power spectrum in a manifestly IR-safe way is 
\begin{align} \label{irsafels}
&P^{\text{IR-safe}}_{ 1-\text{loop}}  \, (a,k)= \frac{D(a)^4}{D(a_i)^4} \int \momspmeas{q} \Big[ p_{13} ( \kvec , \qvec) + p_{22} ( \kvec , \qvec) \Theta_{\rm H} ( | \kvec - \qvec| - q )  \nonumber \\
& \hspace{3in} + p_{22} ( \kvec , -\qvec) \Theta_{\rm H} ( | \kvec + \qvec| - q )  \Big] \ ,
\end{align}
where $\Theta_{\rm H}$ is the Heaviside step function.  Notice that now the only IR divergence is for $q/k \rightarrow 0$ since $p_{22}$ is not integrated near $\qvec \approx \kvec$ any longer (the divergence at $\qvec \rightarrow \kvec$ has been mapped to $q/k \rightarrow 0$), and that the $p_{22}$ term is symmetrized in $\qvec \leftrightarrow - \qvec$, so that the terms proportional to an odd power of $\mu$ explicitly cancel.  Thus, the integrand in \eqn{irsafels} is manifestly IR-safe, in the sense that as $q / k \rightarrow 0$, both the $ d^3 q \, (k/q)^{2} P_{11}(q)$ and $ d^3 q \, (k/q) P_{11} ( q)$ divergences cancel at the level of the integrand.

%
%

\section{IR- and UV-safe integrand with exact time dependence}  \label{exacttime}
In the first part of this section, we would like to find an IR-safe analog of \eqn{irsafels} for $\Lambda$CDM with exact time dependence.\footnote{ \label{footnote:perturbative_exact_time} As discussed in the Introduction, the EdS approximation is known to be correct to better than percent level in $\Lambda$CDM.  Thus, instead of using the exact time dependence, as we do in this paper, one could also expand the time dependence around EdS.  This should be a very good expansion in $\Lambda$CDM, with an expansion parameter of order $1/100$.  Considering the equations of motion \eqn{lineareqs} with $C(a) = 1$, one could do this expansion by setting
\begin{align}
\delta^{(n)} (  a , \kvec ) &= D_+ ( a )^n \left( \bar \delta^{(n)} ( \kvec ) + \epsilon \, \tilde \delta^{(n)} ( a , \kvec ) \right) \\
\Theta^{(n)} ( a , \kvec) & = D_+(a)^n \left( \bar \Theta^{(n)} ( \kvec ) + \epsilon \, \tilde \Theta^{(n)} ( a , \kvec) \right) \ ,
\end{align}
where $\epsilon \sim \mathcal{O}(1/100)$, $\bar \delta^{(n)} ( \kvec )$ and $\bar \Theta^{(n)} ( \kvec )$ are the time-independent EdS fields, and $\tilde \delta^{(n)} ( a , \kvec )$ and $\tilde \Theta^{(n)} ( a , \kvec)$ give the deviations from EdS time dependence.  Plugging this into the equations of motion \eqn{lineareqs}, one can then expand to the desired order in $\epsilon$ (although first order should be sufficient for most purposes).  The result will be differential equations for $\tilde \delta^{(n)} ( a , \kvec )$ and $\tilde \Theta^{(n)} ( a , \kvec)$ which are sourced by $\bar \delta^{(n)} ( \kvec )$ and $\bar \Theta^{(n)} ( \kvec )$.  One can then use the Green's functions of this system to solve for $\tilde \delta^{(n)} ( a , \kvec )$ and $\tilde \Theta^{(n)} ( a , \kvec)$.  The advantage of this approach is that one will in general have less diagrams, because of the linear expansion in $\epsilon$, and there will be less nested time integrals.  Thus, the computation should be much faster.

However, we choose not to pursue this course of action for the following reasons.  First of all, we would like to establish our results with exact time dependence as a matter of principle, and because at one loop it is not too difficult, we choose this path.  Second of all, in this work, in addition to $\Lambda$CDM, we are also interested in the clustering quintessence system for which there is no known analogue of the EdS approximation.  Thus, we leave this expansion for future work.  } We will then construct the IR\&UV-safe integrand in Section~\ref{uvsafesec}, which will be an analgous construction to the one that we present for IR modes.  In doing this, our results can be extended trivially to the adiabatic mode in clustering quintessence described in \cite{Lewandowski:2016yce, Sefusatti:2011cm} by restoring the time-dependent function $C(a)$, which is defined in \eqn{thisisc} (in Appendix~\ref{refform} we summarize the results and notation of the clustering quintessence computation in \cite{Lewandowski:2016yce} for convenience), and is equal to unity for $\Lambda$CDM.  In this paper, our explicit formulae will include the factors of $C(a)$ for completion.  In this paper, we are only interested in the one-loop terms, so we will ignore EFT counterterms (which are trivially IR-safe because they are tree level).

As a side note, we would like to point out that it is only possible to write the IR-safe integrand in clustering quintessence because in the $c_s^2 \rightarrow 0$ limit, quintessence traces dark matter so that the system has only one mode.  In this case, one can always go to the unique freely falling frame of the region to eliminate gradients of the metric and any velocity, thus ensuring IR-safety.  If $c_s^2 \neq 0$, an isocurvature mode is generated, and so we expect the IR cancellation to be spoiled by terms proportional to $c_s^2$.  

\subsection{Expressions for one-loop kernels} \label{oneloopkernelssec}
Below, first we will verify that the leading IR terms of each separate loop cancel, and then we will provide the IR-safe version of the loop integral.  In order to continue, let us write \eqn{P22} and \eqn{P13} (the expressions for $P_{22}$ and $P_{13}$ found in \cite{Lewandowski:2016yce}, which use exact time dependence) in a more useful form:
\begin{align} \label{p22foryou}
P_{22} ( a , k) & = \int \momspmeas{q} \int_0^{a}  d a_2 \int_0^{a_2} d a_1 \, \,  p_{22}(a, a_1 , a_2 ; \kvec , \qvec) \ , \\ 
 P_{13} ( a , k) & = \int \momspmeas{q} \int_0^{a}  d a_2 \int_0^{a_2} d a_1 \, \,  p_{13}(a, a_1 , a_2 ; \kvec , \qvec) \ , \label{p13foryou}
\end{align}
where 
\begin{align} \label{p22expand}
p_{22}(a, a_1 , a_2 ; \kvec , \qvec) & =  \sum_{i = 1}^4 T^{(22)}_i (a, a_1 , a_2) F^{(22)}_i ( \kvec , \qvec) \ ,  \\
p_{13}(a, a_1 , a_2 ; \kvec , \qvec) & =  \sum_{i = 1}^6 T^{(13)}_i (a, a_1 , a_2) F^{(13)}_i ( \kvec , \qvec) \ .\label{p13expand}
\end{align}
We will define the various quantities that enter above momentarily.  First, however, let us comment on the computational strategy.  Each function $p_{22}$ and $p_{13}$ is a sum over terms which are products of a function of time and a function of momentum.  Thus, in order to compute the loop, we must numerically compute the integrals over the functions of time $\int da_2 \int da_1\, T^{(\sigma)}_i$ (where $\sigma \in \{13, 22\}$) separately from the functions of momentum $ \int d^3 q \, F^{(\sigma)}_i$, multiply them, and then add them together.  In order for this approach to be the most numerically efficient, we want \emph{each} of the $F^{(\sigma)}_i ( \kvec , \qvec) $ functions to be IR-safe separately.  That way, numerical uncertainty in the time integrals will not spoil any cancellations that are supposed to happen in the momentum integrals.  Our IR-safe integrand will have this property.   

Next, we will comment on the limits of integration of the time integrals.  Notice that in \eqn{P22} for $P_{22}$, the limits are $ \int_0^{a}  d a_2 \int_0^{a} d a_1$, while in \eqn{P13} for $P_{13}$ they are $ \int_0^{a}  d a_2 \int_0^{a_2} d a_1$.  In order to compare the $P_{13}$ and $P_{22}$ integrands directly, we want them to have the same limits, so we use the following fact, valid for any function $f(a_1 , a_2)$, 
\begin{align}
\int_0^{a}  d a_2 \int_0^{a} d a_1  \, f( a_1 , a_2) & = \left( \int_0^{a}  d a_2 \int_0^{a_2} d a_1 + \int_0^a d a_1 \int_0^{a_1} d a_2\right) f(a_1 , a_2) \\
& = \int_0^{a}  d a_2 \int_0^{a_2} d a_1 \left( f( a_1 , a_2) + f( a_2 , a_1) \right) \ , 
\end{align}
where in the first passage we re-parametrize the square region $0 \leq a_1 \leq a$ and $0 \leq a_2 \leq a$, and in the second passage we redefined the variables of integration in the second term.  This allows us to write \eqn{p22foryou} and \eqn{p13foryou}, in a way in which $P_{22}$ and $P_{13}$ have the same limits for the time integrals.

{Let us now go back to defining the terms that appear in \eqn{p22expand} and \eqn{p13expand}.}  In order to write the time dependent coefficients $T^{(22)}_i$ and $T^{(13)}_i$, we first define various $\bar G$, the part of the Green's functions \eqn{gdelta} - \eqn{gtheta} that do not contain the Heaviside function, as 
\be
G^{\delta, \Theta}_{1,2}( a_1 , a_2 ) \equiv \bar G^{\delta,\Theta}_{1,2}( a_1 , a_2 ) \, \Theta_{\rm H}( a_1 - a_2) \ . 
\ee
Using \eqn{2nd-solution} - \eqn{2nd-solution2} and \eqn{timedepcoeff1} - \eqn{timedepcoeff2}, this leads us to write the time dependent coefficients $T^{(13)}_i$ as (first making the replacement $f_+ (a ) = a \,   D'_+(a) / D_+(a)$ everywhere, where $D'_+ (a) \equiv \partial_a D_+ (a)$)
\begin{align}
T^{(13)}_1 ( a , a_1 , a_2) & = K( a , a_1 , a_2) \,  \bar G^\delta_1 ( a , a_2) \bar G^\delta_1 ( a_2 , a_1)  \ ,\\
T^{(13)}_2 ( a , a_1 , a_2) & = K( a , a_1 , a_2) \,  \bar G^\delta_1 ( a , a_2) \bar G^\delta_2 ( a_2 , a_1) \ , \\
T^{(13)}_3 ( a , a_1 , a_2) & = K( a , a_1 , a_2) \,  \bar G^\delta_1 ( a , a_2) \bar G^\Theta_1 ( a_2 , a_1) \ ,\\
T^{(13)}_4 ( a , a_1 , a_2) & = K( a , a_1 , a_2) \,  \bar G^\delta_1 ( a , a_2) \bar G^\Theta_2 ( a_2 , a_1) \ ,\\
T^{(13)}_5 ( a , a_1 , a_2) & = K( a , a_1 , a_2) \,  \bar G^\delta_2 ( a , a_2) \bar G^\Theta_1 ( a_2 , a_1)\ ,\\
T^{(13)}_6 ( a , a_1 , a_2) & = K( a , a_1 , a_2) \,  \bar G^\delta_2 ( a , a_2) \bar G^\Theta_2 ( a_2 , a_1) \ ,
\end{align}
where the common factor $K$ is given by 
\be
K( a , a_1 , a_2 ) = \frac{ a_1 a_2 D_+(a ) D_+ ( a_1) D'_+(a_1) D'_+(a_2)}{C(a_1) C(a_2) D_+(a_i)^4} \ .
\ee
The other time-dependent coefficients $T^{(22)}_i$ are given by 
\begin{align}
T^{(22)}_1 ( a , a_1 , a_2) &= K( a_2 , a_1 , a_2)  \, 2\,   \bar G^\delta_1 ( a , a_1) \bar G^\delta_1 ( a , a_2) \ ,\\
T^{(22)}_2 ( a , a_1 , a_2) &= K( a_2 , a_1 , a_2)  \, \left(    \bar G^\delta_1 ( a , a_1) \bar G^\delta_2 ( a , a_2)  + \bar G^\delta_2 ( a , a_1) \bar G^\delta_1 ( a , a_2)  \right) \ , \\
T^{(22)}_3 ( a , a_1 , a_2) &=  T^{(22)}_2 ( a , a_1 , a_2) \ ,\\
T^{(22)}_4 ( a , a_1 , a_2) & = K( a_2 , a_1 , a_2)  \, 2\,   \bar G^\delta_2 ( a , a_1) \bar G^\delta_2 ( a , a_2) \ . 
\end{align}

With the above definition of the time dependent $T^{(\sigma)}_i$ functions, we have the following momentum dependent functions:
\begin{align} \label{longlist1}
F^{(13)}_1 ( \kvec , \qvec) & = 4 \,  \alpha_s ( \kvec , \qvec) \, \alpha ( - \qvec , \kvec + \qvec) \, P^{\rm in}_{\kvec}\, P^{\rm in}_{\qvec} \ , \\
F^{(13)}_2 ( \kvec , \qvec) & = 4 \,  \beta ( \kvec , \qvec) \, \alpha ( - \qvec , \kvec + \qvec) \, P^{\rm in}_{\kvec}\, P^{\rm in}_{\qvec} \ , \\ 
F^{(13)}_3 ( \kvec , \qvec) & = 4 \,  \alpha_s ( \kvec , \qvec) \, \alpha ( \kvec + \qvec , - \qvec)\, P^{\rm in}_{\kvec}\, P^{\rm in}_{\qvec}  \ , \\
F^{(13)}_4 ( \kvec , \qvec) & = 4 \,  \beta ( \kvec , \qvec) \, \alpha (  \kvec + \qvec, -\qvec) \, P^{\rm in}_{\kvec}\, P^{\rm in}_{\qvec} \ , \\
F^{(13)}_5 ( \kvec , \qvec) & =  4 \times 2 \,  \alpha_s ( \kvec , \qvec) \, \beta ( - \qvec , \kvec + \qvec)\, P^{\rm in}_{\kvec}\, P^{\rm in}_{\qvec} \ , \\
F^{(13)}_6 ( \kvec , \qvec) & = 4 \times 2 \,  \beta ( \kvec , \qvec) \, \beta ( - \qvec , \kvec + \qvec) \, P^{\rm in}_{\kvec}\, P^{\rm in}_{\qvec} \ , \label{longlist4}
\end{align}
and 
\begin{align} \label{longlist3}
F^{(22)}_1 ( \kvec , \qvec ) & = 2 \, \alpha_s( \kvec - \qvec , \qvec )^2  \, P^{\rm in}_{\kvec - \qvec}\, P^{\rm in}_{\qvec} \ ,   \\
F^{(22)}_2 ( \kvec , \qvec ) & = 2 \, \alpha_s( \kvec - \qvec , \qvec ) \, \beta ( \kvec - \qvec , \qvec) \, P^{\rm in}_{\kvec - \qvec}\, P^{\rm in}_{\qvec} \ ,\\
F^{(22)}_3 ( \kvec , \qvec )  & = 2 \, \alpha_s( \kvec - \qvec , \qvec ) \, \beta ( \kvec - \qvec , \qvec)  \, P^{\rm in}_{\kvec - \qvec}\, P^{\rm in}_{\qvec} \ ,\\
F^{(22)}_4 ( \kvec , \qvec ) & = 2 \, \beta( \kvec - \qvec , \qvec ) \, \beta ( \kvec - \qvec , \qvec) \, P^{\rm in}_{\kvec - \qvec}\, P^{\rm in}_{\qvec} \ ,  \label{longlist2}
\end{align}
where in the above $\alpha$ and $\beta$ are the standard interaction functions from dark-matter perturbation theory
\begin{align}  
\alpha ( \qvec_1 , \qvec_2 ) & = 1 + \frac{\qvec_1 \cdot \qvec_2}{q_1^2}  \ , \label{alphadef2} \\
\beta( \qvec_1 , \qvec_2 ) & = \frac{ | \qvec_1 + \qvec_2 |^2 \qvec_1 \cdot \qvec_2}{2 q_1^2 q_2^2}    \label{betadef2} \ ,
 \end{align}
and $\alpha_s ( \qvec_1 , \qvec_2) = \half ( \alpha( \qvec_1 , \qvec_2 ) + \alpha ( \qvec_2 , \qvec_1) )$.  To get the compact forms in \eqn{longlist1} - \eqn{longlist2}, we have used the properties that $\alpha_s$ and $\beta$ are symmetric, that $\alpha( \qvec_1 , - \qvec_2) = \alpha(- \qvec_1 , \qvec_2)$ and $\beta( \qvec_1 , - \qvec_2) = \beta(- \qvec_1 , \qvec_2)$, and switched the variable of integration from $\qvec$ to $ - \qvec $ in some terms.


 \subsection{IR-limit} \label{irlimitsec}
 In this section, we will examine the IR properties of the integrands $p_{22}$ and $p_{13}$ and show that both the leading and subleading IR divergences must cancel when the full one-loop contribution is computed.  In the next section we will use the IR limits found here to write the manifestly IR-safe integrand.  In $p_{13}$, the only IR divergence is for $q / k \rightarrow 0$, for which we have the following limits:
 \begin{align}
 F^{(13)}_{1} ( \kvec , \qvec) & =  - 2 \mu^2 \left( \frac{k^2}{q^2} +1  \right) P^{\rm in}_{\kvec} P^{\rm in}_{\qvec} \ ,\\
  F^{(13)}_{2} ( \kvec , \qvec) & =  - 2 \mu^2 \left( \frac{k^2}{q^2} + 1 \right) P^{\rm in}_{\kvec} P^{\rm in}_{\qvec} \ , \\
  F^{(13)}_{3} ( \kvec , \qvec) & \rightarrow  -2 \left( -2 + \mu^2 + \mathcal{O}\left(\frac{q^2}{k^2}\right)   \right)    P^{\rm in}_{\kvec} P^{\rm in}_{\qvec} \ , \\ 
  F^{(13)}_{4} ( \kvec , \qvec) &  = 2 \mu^2  P^{\rm in}_{\kvec} P^{\rm in}_{\qvec} \ , \\
  F^{(13)}_{5} ( \kvec , \qvec) &  \rightarrow - 2  \left( \mu^2 \frac{k^2}{q^2} +2 - 6 \mu^2 +4 \mu^4  + \mathcal{O}\left(\frac{q^2}{k^2}\right) \right) P^{\rm in}_{\kvec} P^{\rm in}_{\qvec} \ ,\\
    F^{(13)}_{6} ( \kvec , \qvec) & =  - 2 \mu^2  \frac{k^2}{q^2} P^{\rm in}_{\kvec} P^{\rm in}_{\qvec} \ , 
 \end{align}
 where in the above, relations with an $``="$ sign are exact relations, independent from the limit $q / k \rightarrow 0$, and relations with a $``\rightarrow"$ sign are only valid in the limit $q / k \rightarrow 0$.  We are concerned with the IR terms that we know must cancel because of the equivalence principle, i.e. the ones proportional to $k^2 / q^2$ and $k/q$, so we define
 \begin{align} \label{f1ir}
   F^{(13)}_{1 , \text{IR}} ( \kvec , \qvec) &=  F^{(13)}_{2, \text{IR}} ( \kvec , \qvec) =  F^{(13)}_{5, \text{IR}} ( \kvec , \qvec) =   F^{(13)}_{6, \text{IR}} ( \kvec , \qvec)  =  - 2 \mu^2  \frac{k^2}{q^2} P^{\rm in}_{\kvec} P^{\rm in}_{\qvec}\\
 F^{(13)}_{3 , \text{IR}} ( \kvec , \qvec) &=  F^{(13)}_{4, \text{IR}} ( \kvec , \qvec) = 0  \, \label{f3ir}
 \end{align}
so that $F^{(13)}_{i,\text{IR}} ( \kvec, \qvec)= \lim_{q/k \rightarrow 0} F^{(13)}_{i} ( \kvec, \qvec)$ to order $k / q$.  
 
 The analysis of the $p_{22}$ integrand is slightly more complicated because there are two IR divergences: one for $q / k \rightarrow 0$, and one for $\qvec \rightarrow \kvec$.  However, these divergences are really the same, since $p_{22} ( a , a_1 , a_2; \kvec ,  \qvec) = p_{22} ( a , a_1 , a_2; \kvec , \kvec - \qvec)$, as can be seen directly in \eqn{longlist3} - \eqn{longlist2}.  In the next section, we will see that this allows us to write the IR-safe integrand such that the only IR divergence is for $q / k \rightarrow 0$, so we will provide that limit here.  Thus, for $q/k \rightarrow 0$, we have 
 \begin{align} \label{firstf22}
 F^{(22)}_1 ( \kvec , \qvec ) & \rightarrow  \left( \frac{ \mu^2 }{2}  \frac{k^2}{q^2}  +\frac{k}{q} \left[  \mu  - \frac{\mu^3}{2}   \frac{\partial \log P^{\rm in}_{\kvec}}{ \partial \log k}\right]  + \mathcal{O}\left( \frac{k^0}{q^0} \right) \right) P^{\rm in}_{\kvec} P^{\rm in}_{\qvec} \ , \\
  F^{(22)}_2 ( \kvec , \qvec ) & \rightarrow  \left( \frac{ \mu^2 }{2}  \frac{k^2}{q^2}  +\frac{k}{q} \left[  \mu^3  - \frac{\mu^3}{2}   \frac{\partial \log P^{\rm in}_{\kvec}}{ \partial \log k}\right]  + \mathcal{O}\left( \frac{k^0}{q^0} \right) \right) P^{\rm in}_{\kvec} P^{\rm in}_{\qvec}  \ , \\
  F^{(22)}_3 ( \kvec , \qvec ) & \rightarrow  \left( \frac{ \mu^2 }{2}  \frac{k^2}{q^2}  +\frac{k}{q} \left[  \mu^3  - \frac{\mu^3}{2}   \frac{\partial \log P^{\rm in}_{\kvec}}{ \partial \log k}\right]  + \mathcal{O}\left( \frac{k^0}{q^0} \right) \right) P^{\rm in}_{\kvec} P^{\rm in}_{\qvec} \ , \\
    F^{(22)}_4 ( \kvec , \qvec ) & \rightarrow  \left( \frac{ \mu^2 }{2}  \frac{k^2}{q^2}  +\frac{k}{q} \left[ - \mu + 2 \mu^3  - \frac{\mu^3}{2}   \frac{\partial \log P^{\rm in}_{\kvec}}{ \partial \log k}\right]  + \mathcal{O}\left( \frac{k^0}{q^0} \right) \right) P^{\rm in}_{\kvec} P^{\rm in}_{\qvec} \ . \label{lastf22}
 \end{align}
 From this, we see that the only divergences proportional to $k/q$ comes from $p_{22}$, but that all of those terms are proportional to an odd power of $\mu$.  Thus, we can eliminate them at the level of the integrand by using a quantity that is manifestly even for $\qvec \rightarrow - \qvec$, which we do below in \eqn{newp22symm}.  For now, similar to \eqn{f1ir} and \eqn{f3ir}, we can define the IR divergent terms of $p_{22}$ as
 \begin{align} \label{firstf22ir}
 F^{(22)}_{1,{\rm IR}} ( \kvec , \qvec ) & =  \left( \frac{ \mu^2 }{2}  \frac{k^2}{q^2}  +\frac{k}{q} \left[  \mu  - \frac{\mu^3}{2}   \frac{\partial \log P^{\rm in}_{\kvec}}{ \partial \log k}\right]  \right) P^{\rm in}_{\kvec} P^{\rm in}_{\qvec} \ , \\
  F^{(22)}_{2,{\rm IR}} ( \kvec , \qvec ) & =  \left( \frac{ \mu^2 }{2}  \frac{k^2}{q^2}  +\frac{k}{q} \left[  \mu^3  - \frac{\mu^3}{2}   \frac{\partial \log P^{\rm in}_{\kvec}}{ \partial \log k}\right]   \right) P^{\rm in}_{\kvec} P^{\rm in}_{\qvec} \ ,  \\
  F^{(22)}_{3,{\rm IR}} ( \kvec , \qvec ) & =  \left( \frac{ \mu^2 }{2}  \frac{k^2}{q^2}  +\frac{k}{q} \left[  \mu^3  - \frac{\mu^3}{2}   \frac{\partial \log P^{\rm in}_{\kvec}}{ \partial \log k}\right] \right) P^{\rm in}_{\kvec} P^{\rm in}_{\qvec} \ ,\\
    F^{(22)}_{4,{\rm IR}} ( \kvec , \qvec ) & =  \left( \frac{ \mu^2 }{2}  \frac{k^2}{q^2}  +\frac{k}{q} \left[ - \mu + 2 \mu^3  - \frac{\mu^3}{2}   \frac{\partial \log P^{\rm in}_{\kvec}}{ \partial \log k}\right]   \right) P^{\rm in}_{\kvec} P^{\rm in}_{\qvec} \ , \label{lastf22ir}
 \end{align}
 so that $F^{(22)}_{i,\text{IR}} ( \kvec, \qvec)= \lim_{q/k \rightarrow 0} F^{(22)}_{i} ( \kvec, \qvec)$ to order $k / q$.


 \subsection{IR-safe integrand} \label{irsafeintegrandsec}
 In this section, we write the manifestly IR-safe version of $P_{1-\text{loop}}$, in such a way that the leading IR divergences in \eqn{p22foryou} and \eqn{p13foryou}, proportional to $k^2/q^2$ and $k/q$, which must cancel, are absent at the level of the integrand.  First, we verify that the leading IR terms indeed cancel (this exercise will also be useful in defining the IR-safe integrand anyway).  To start, we can manipulate $P_{22}$ analogously to \cite{Carrasco:2013sva} to get 
 \begin{align} \nonumber
& P_{22}( a , k)  = \int \momspmeas{q} \int_0^{a}  d a_2 \int_0^{a_2} d a_1 \, \,  \Big( p_{22}(a, a_1 , a_2 ; \kvec , \qvec)\,  \Theta_{\rm H}( |\kvec - \qvec| - q ) \\
&  \hspace{3in} + p_{22} ( a , a_1 , a_2; \kvec , - \qvec)\, \Theta_{\rm H}( | \kvec + \qvec| - q ) \Big) \label{newp22symm}
\end{align}
where we have used the fact that $p_{22} ( a , a_1 , a_2; \kvec ,  \qvec) = p_{22} ( a , a_1 , a_2; \kvec , \kvec - \qvec)$, as discussed above.  This expression has two advantages.  First, the subleading divergence proportional to $k/q$ manifestly cancels, because all of those terms were odd in $\qvec$ as discussed above.  Second, the divergence in $p_{22}$ for $\qvec \rightarrow \kvec$ has been mapped to $q/k \rightarrow 0$, and the integral no longer involves the region $\qvec \approx \kvec$.  Then, we can write the full one-loop contribution as 
\begin{align}\nonumber
& P_{1-\text{loop}} ( a , k ) =  \int \momspmeas{q} \int_0^{a}  d a_2 \int_0^{a_2} d a_1 \Big( p_{13} ( a , a_1 , a_2; \kvec ,  \qvec) +  p_{22}(a, a_1 , a_2 ; \kvec , \qvec)\,  \Theta_{\rm H}( |\kvec - \qvec| - q ) \\ 
&  \hspace{3in} + p_{22} ( a , a_1 , a_2; \kvec , - \qvec)\, \Theta_{\rm H}( | \kvec + \qvec| - q ) \Big) \ .  \label{P1loop1}
\end{align}
We already know that the term proportional to $k/q$ cancels, so we would like to verify now that the leading IR parts (which only come from $q / k \rightarrow 0$ now because $p_{22}$ is not integrated near $\qvec \approx \kvec$ any longer) of the integrand in \eqn{P1loop1} cancel.

The equation \eqn{P1loop1} is the start to finding the IR-safe integrand, but it will also help us explicitly verify that the IR divergences cancel.  In order to check the cancellation, we consider the integrand as $q / k \rightarrow 0$.  In that limit, both of the Heaviside functions can be taken to be unity, so we need to consider the $k^2 / q^2$ terms of (as we said, the $k/q$ terms cancel already)
\be
p_{13} ( a , a_1 , a_2; \kvec ,  \qvec) + 2 \, p_{22} ( a , a_1 , a_2; \kvec ,  \qvec) \ . 
\ee
These are 
\begin{align} \nonumber
& \frac{p_{13} ( a , a_1 , a_2; \kvec ,  \qvec) }{P^{\rm in}_{\kvec} P^{\rm in}_{\qvec}} \rightarrow -2 \mu^2 \frac{k^2}{q^2} K( a , a_1 , a_2) \Big(   \bar G^\delta_1 ( a , a_2) \bar G^\delta_1 ( a_2 , a_1) +  \bar G^\delta_1 ( a , a_2) \bar G^\delta_2 ( a_2 , a_1)  \\
& \hspace{2.7in} +  \bar G^\delta_2 ( a , a_2) \bar G^\Theta_1 ( a_2 , a_1) +  \bar G^\delta_2 ( a , a_2) \bar G^\Theta_2 ( a_2 , a_1) \Big) \label{p13irlimit}
\end{align}
and
\begin{align}
& \frac{2 \, p_{22} ( a , a_1 , a_2; \kvec ,  \qvec) }{P^{\rm in}_{\kvec} P^{\rm in}_{\qvec}} \rightarrow 2 \mu^2 \frac{k^2}{q^2} K( a_2 , a_1 , a_2) \Big(   \bar G^\delta_1 ( a , a_1) \bar G^\delta_1 ( a , a_2) +  \bar G^\delta_1 ( a , a_1) \bar G^\delta_2 ( a , a_2) \nonumber \\
& \hspace{2.7in} +  \bar G^\delta_2 ( a , a_1) \bar G^\delta_1 ( a , a_2) +  \bar G^\delta_2 ( a , a_1) \bar G^\delta_2 ( a , a_2) \Big) \ .  \label{p22irlimit}
\end{align}
Thus, these two will cancel if  
\begin{align} \nonumber
& D_+( a ) \left(  \bar G^\delta_1 ( a , a_2) \bar G^\delta_1 ( a_2 , a_1) +  \bar G^\delta_1 ( a , a_2) \bar G^\delta_2 ( a_2 , a_1) +   \bar G^\delta_2 ( a , a_2) \bar G^\Theta_1 ( a_2 , a_1) +  \bar G^\delta_2 ( a , a_2) \bar G^\Theta_2 ( a_2 , a_1)  \right)  \\
& + D_+( a_2) \left(  \bar G^\delta_1 ( a , a_1) \bar G^\delta_1 ( a , a_2) +  \bar G^\delta_1 ( a , a_1) \bar G^\delta_2 ( a , a_2)  +  \bar G^\delta_2 ( a , a_1) \bar G^\delta_1 ( a , a_2) +  \bar G^\delta_2 ( a , a_1) \bar G^\delta_2 ( a , a_2)     \right) \nonumber \\ 
& \hspace{.1in} = 0  \label{gfidentity}
\end{align}
and indeed, one can check that this is the case by using the explicit expressions \eqn{gdelta} - \eqn{gtheta}.  Thus, we have successfully shown that the IR divergences in $P_{1-\text{loop}}$ (proportional to $k^2 / q^2$ and $k/q$ respectively) cancel.  In particular, while $P_{13}$ and $P_{22}$ have different time dependences, this shows that the leading IR terms in fact have the same time dependence, and that this allows these terms to cancel.
 
Now, we notice that at finite $q$ each diagram has a different time dependence, which becomes the same only in the limit $q/k \rightarrow 0$. This has the unfortunate consequence that the cancellation in the IR will happen only if the time integrals are computed very accurately. This inconvenience can be avoided by doing the following procedure. We can first add and subtract out of each diagram the IR divergent part. The sum of all the divergences of a single diagram combine themselves into a term that has some given time dependence times a common momentum dependent factor. This time dependence is the same as the one associated to the sum of the IR divergencies of the other diagram, with the same momentum dependent factor.  A relative minus sign ensures the cancellation.  This manipulation guarantees that the IR divergent terms never enter the computation at all, thus making each separate contribution IR-safe.  This is necessary, contrary to the case studied in \cite{Carrasco:2013sva}, because we are summing together many different contributions which are products of integrals over time and integrals over momentum.  Thus, in order for the final answer to be the most computationally efficient, \emph{each} of the contributions must be manifestly IR-safe.  Concretely, our procedure is the following.  We start with the integrand of $P_{1-\text{loop}}$ in \eqn{P1loop1}:
\begin{align}
 p_{1-\text{loop}} ( a , a_1 , a_2; \kvec ,  \qvec) \equiv p_{13} ( a , a_1 , a_2; \kvec ,  \qvec) &+  p_{22}(a, a_1 , a_2 ; \kvec , \qvec)\,  \Theta_{\rm H}( |\kvec - \qvec| - q )  \nonumber\\ 
 & +  p_{22}(a, a_1 , a_2 ; \kvec , - \qvec)\,  \Theta_{\rm H}( |\kvec + \qvec| - q ) \ .
\end{align}
To this, we will momentarily add and subtract the IR terms through the function 
\begin{align}
 p_{1-\text{loop}}^{\text{IR}} ( a , a_1 , a_2; \kvec ,  \qvec) \equiv p_{13}^{\text{IR}} ( a , a_1 , a_2; \kvec ,  \qvec) & + p_{22}^{\text{IR}} ( a , a_1 , a_2; \kvec ,  \qvec) \nonumber \\
& + p_{22}^{\text{IR}} ( a , a_1 , a_2; \kvec , - \qvec)  
\end{align}
where 
\begin{align}
 p_{13}^{\text{IR}} ( a , a_1 , a_2; \kvec ,  \qvec)  & =  \sum_{i=1}^6 T^{(13)}_i ( a , a_1 , a_2)  F^{(13)}_{i , \text{IR}}(\kvec , \qvec) \Theta_{\rm H} ( k - q)  \ , \\
 p_{22}^{\text{IR}} ( a , a_1 , a_2; \kvec ,  \qvec)  & =  \sum_{i=1}^4 T^{(22)}_i ( a , a_1 , a_2)  F^{(22)}_{i , \text{IR}}(\kvec , \qvec)   \Theta_{\rm H} ( k - q) \ .
 \end{align}
The functions $ p_{13}^{\text{IR}} $ and $ p_{22}^{\text{IR}} $ are nothing but the original expressions for $p_{13}$ and $p_{22}$ from \eqn{p13expand} and \eqn{p22expand}, but with $F^{(\sigma)}_i$ replaced by $F^{(\sigma)}_{i,\text{IR}}$ and multiplied by a Heaviside function $\Theta_{\rm H} ( k - q)$ so that the UV is unchanged.  This means that we can express the integrand of $P_{1-\text{loop}}$ as
\begin{align}
 p_{1-\text{loop}}( a , a_1 , a_2; \kvec ,  \qvec) & =   \left( p_{1-\text{loop}}( a , a_1 , a_2; \kvec ,  \qvec) -  p_{1-\text{loop}}^{\text{IR}}( a , a_1 , a_2; \kvec ,  \qvec) \right) \nonumber \\
 & \hspace{.5in} + p_{1-\text{loop}}^{\text{IR}}( a , a_1 , a_2; \kvec ,  \qvec) \label{irsubtract}
 \end{align}
by simply adding and subtracting the IR parts, so that overall the integrand is unchanged.  

Let us now examine separately the terms inside and outside of the parentheses on the right-hand side of \eqn{irsubtract}.  We start with the terms inside of the parentheses and rearrange them so that we subtract the IR terms from each contribution to $p_{1-\text{loop}}$ individually to get 
\begin{align}
& p_{1-\text{loop}}( a , a_1 , a_2; \kvec ,  \qvec) -  p_{1-\text{loop}}^{\text{IR}}( a , a_1 , a_2; \kvec ,  \qvec)   = \nonumber\\
& \hspace{2in}     p_{13}^{\text{IR-safe}} ( a , a_1 , a_2; \kvec ,  \qvec) +   p_{22}^{\text{IR-safe}} (a, a_1 , a_2 ; \kvec , \qvec)   \ ,
\end{align}
where 
\begin{align} \label{deltap13}
 p_{13}^{\text{IR-safe}} ( a , a_1 , a_2; \kvec ,  \qvec)  &= \sum_{i=1}^6 T^{(13)}_i ( a , a_1 , a_2) \,  F^{(13)}_{i,\text{IR-safe}}(\kvec , \qvec) \ , \\
 p_{22}^{\text{IR-safe}} ( a , a_1 , a_2; \kvec ,  \qvec) & = \sum_{i=1}^4 T^{(22)}_i ( a , a_1 , a_2) \Big(  F^{(22)}_{i,\text{IR-safe}}(\kvec , \qvec) +   F^{(22)}_{i,\text{IR-safe}}(\kvec , - \qvec) \Big) \ , \label{deltap22}
\end{align}
and 
 \begin{align}
  F^{(13)}_{i,\text{IR-safe}} (\kvec , \qvec)  & = F^{(13)}_i(\kvec , \qvec) -  F^{(13)}_{i , \text{IR}}(\kvec , \qvec) \, \Theta_{\rm H} ( k - q) \ , \\
  F^{(22)}_{i,\text{IR-safe}} (\kvec , \qvec) & = F^{(22)}_i(\kvec , \qvec) \, \Theta_{\rm H} ( | \kvec - \qvec| - q )  -  F^{(22)}_{i , \text{IR}}(\kvec , \qvec) \, \Theta_{\rm H} ( k - q) \ . 
 \end{align} 
By definition, each of the $ F^{(22)}_{i,\text{IR-safe}}$ and $ F^{(13)}_{i,\text{IR-safe}}$ integrands are IR-safe because we have subtracted the IR divergences explicitly.

Now let us look at the other term on the right-hand side of \eqn{irsubtract} which is 
\begin{align} \label{thisiszero}
 p_{1-\text{loop}}^{\text{IR}}( a , a_1 , a_2; \kvec ,  \qvec) =  &\sum_{i=1}^6 T^{(13)}_i ( a , a_1 , a_2) \, F^{(13)}_{i , \text{IR}}(\kvec , \qvec)  \, \Theta_{\rm H} ( k - q)   \\
 & +  \sum_{i=1}^4 T^{(22)}_i ( a , a_1 , a_2) \, \left( F^{(22)}_{i , \text{IR}}(\kvec , \qvec) + F^{(22)}_{i , \text{IR}}(\kvec , -\qvec) \right)  \, \Theta_{\rm H} ( k - q)   = 0 \ ,   \nonumber
\end{align}
and is zero simply because the IR divergences cancel (as we have already shown), i.e. it follows from \eqn{p13irlimit}, \eqn{p22irlimit}, and \eqn{gfidentity}.  Thus, this term does not contribute at all to the one-loop integral, and we are finally left with\footnote{\label{footnotef} We would like to comment that instead of doing the manipulation that maps the IR divergence in $P_{22}$ from $\qvec \rightarrow \kvec$ to $q/k \rightarrow 0$, one could directly subtract out the divergence at $\qvec \rightarrow \kvec$ at the level of the integrand.  This more straightforward approach could be advantageous for computations higher than one loop because the momentum dependence of the integrands becomes more complicated as one integrates over more and more loop momenta.}  
\begin{align}
& P^{\text{IR-safe}}_{ 1-\text{loop}}  ( a , k ) =  \int \momspmeas{q} \int_0^{a}  d a_2 \int_0^{a_2} d a_1 \Big(  p_{13}^{\text{IR-safe}} ( a , a_1 , a_2; \kvec ,  \qvec) +   p_{22}^{\text{IR-safe}}(a, a_1 , a_2 ; \kvec , \qvec)  \Big) \ ,  \label{P1loop2}
\end{align} 
with $ p_{13}^{\text{IR-safe}}$ and $  p_{22}^{\text{IR-safe}}$ given in \eqn{deltap13} and \eqn{deltap22}.  Of course, $P^{\text{IR-safe}}_{ 1-\text{loop}}  ( a , k ) = P_{1-\text{loop}} ( a , k)$ because we have simply added zero to $P_{1-\text{loop}}$, but we put the extra label ``IR-safe" to denote that it is computed with the IR-safe integrand.
Thus, we have arrived at our final expressions \eqn{P1loop2}, \eqn{deltap13} and \eqn{deltap22}.  We have shown that the IR divergences (proportional to $k^2 / q^2$ and $k/q$) in the one-loop, equal-time power spectrum with exact time dependence in $\Lambda$CDM cancel (as was required by the equivalence principle), and we have provided integrands which are IR-safe during every step of the numerical integration.  By keeping the factor of $C(a)$ in the various time-dependent coefficients, our results trivially extend to the adiabatic mode in clustering quintessence.

Before we end this discussion of IR effects, let us briefly comment on the very important task of correctly describing the baryon acoustic oscillations (BAO).  
  The IR-resummation is a way to controllably include the effects of long-wavelength displacements on the BAO peak. The relevant formulae for the various systems (dark matter, galaxies, and redshift space) were developed in~\cite{Senatore:2014via, Angulo:2014tfa, Senatore:2014vja, Lewandowski:2015ziq, Perko:2016puo}. Ref.~\cite{Baldauf:2015xfa} provided a simplification (with approximations) of the same formulas applied to dark matter in real space.  The IR-resummation can be applied directly to the exact time-dependence power spectra presented in this paper, using the standard formulae presented in previous works for the resummation.  This is because the difference between the exact time-dependence displacements and the EdS approximated displacements is very small (indeed it is zero in the limit that we treat the displacements as linear).  Moreover, this difference can be recovered order by order in the perturbative expansion of the resummation formula.  In other words, one can use the approximate resummation matrix which uses EdS displacements (call it $M_{\rm EdS}$), and the formulae will automatically recover the correct resummation as a Taylor expansion in $M_{\rm true} - M_{\rm EdS}$. This is a remarkable property of the formula developed in~\cite{Senatore:2014via}.

\subsection{UV-safe integrand} \label{uvsafesec}

In addition to the spurious IR terms which we have shown must cancel in the final expression for $\Poneloop$, there are also spurious UV contributions to the individual momentum-dependent functions $F^{(\sigma)}_i$ that must cancel in the full one-loop result.  These spurious UV divergences are not present with the approximate time dependence, because in that case each diagram has a common time factor, and the structure of the UV divergences (which are different for each diagram) forces them to cancel automatically.

The same reasoning applies to the UV terms as to the IR terms: full cancellation only happens when the time dependent coefficients are determined with infinite precision, so a mistake in the time integrals can cause the large UV terms to not cancel completely.  This means that one has to run the integrals with much more precision than the precision desired in the final answer.  To address this, in this section we will provide a manifestly UV-safe integrand using a procedure directly analogous to the one we used for the IR terms: we will subtract out the spurious UV terms at the level of the integrand so that they never enter the computation at all.  As in the presentation of the IR-safe integrand, the exact time dependence makes the cancellation more opaque, but we show how it happens below.  Additionally, one could subtract the UV terms of the loop integrals which are degenerate with the counterterms (discussed more below).  As noted in \cite{Foreman:2015lca} this has the advantage that one does not waste computational time computing a (large) part of the loop which will ultimately be adjusted by the counterterms, and so less numerical precision can be used on the loop computation to obtain a desired final precision.  

First, let us consider, in general, the UV dependence of the individual loop terms.  There can be no terms at lower order in $k/q$ for $k/q \rightarrow 0 $ than
\begin{align}
P_{13}^{\rm UV} ( k ) & \propto  k^2 P_{11}(k) \int \momspmeas{q}  \frac{1}{q^2} P_{11}(q) \ ,  \\ 
P_{22}^{\rm UV} ( k ) & \propto k^4 \int \momspmeas{q} \frac{1}{q^4} P_{11} ( q ) P_{11} ( q ) \ .
\end{align}
Notice that these are just the terms of the power spectrum that can be adjusted by counterterms: the $k^2 P_{11} ( k )$ counterterm comes from the $\partial^2 \delta$ term in the stress tensor, and the $k^4$ term comes from the stochastic counterterm \cite{Carrasco:2012cv}.  In principle, one does not even need to compute these terms in the loop, since they are degenerate with the counterterms~\cite{Foreman:2015lca}.  For simplicity, in this paper, we choose not to cancel at the integrand level the pieces that are degenerate with counterterms, although the procedure to do this is a straightforward extension of what we present below, where we focus on cancelling only the UV divergences that cannot be cancelled by a counterterm and so must cancel at the level of the integrand.  

Let us now find the UV terms in our integrands in \eqn{longlist1}~-~\eqn{longlist4} and \eqn{longlist3}~-~\eqn{longlist2}.  For $k/q \rightarrow 0$, we have 
 \begin{align}
 F^{(13)}_{1} ( \kvec , \qvec) & =  - 2 \mu^2 \left( 1+ \frac{k^2}{q^2}   \right) P^{\rm in}_{\kvec} P^{\rm in}_{\qvec} \ ,\\
  F^{(13)}_{2} ( \kvec , \qvec) & =  - 2 \mu^2 \left( 1+ \frac{k^2}{q^2} \right) P^{\rm in}_{\kvec} P^{\rm in}_{\qvec} \ , \\
  F^{(13)}_{3} ( \kvec , \qvec) & \rightarrow  2 \left(  \mu^2 + \frac{k^2}{q^2} \left( 2 - 6 \mu^2 +4 \mu^4\right) + \mathcal{O}\left( \frac{k^4}{q^4} \right)     \right)     P^{\rm in}_{\kvec} P^{\rm in}_{\qvec} \ , \\ 
  F^{(13)}_{4} ( \kvec , \qvec) &  = 2 \mu^2  P^{\rm in}_{\kvec} P^{\rm in}_{\qvec} \ , \\
  F^{(13)}_{5} ( \kvec , \qvec) &  \rightarrow \left(  \frac{k^2}{q^2} \left(  - 4 + 2 \mu^2  \right) + \mathcal{O} \left( \frac{k^4}{q^4} \right)    \right)  P^{\rm in}_{\kvec} P^{\rm in}_{\qvec} \ ,\\
    F^{(13)}_{6} ( \kvec , \qvec) & =  - 2 \mu^2  \frac{k^2}{q^2} P^{\rm in}_{\kvec} P^{\rm in}_{\qvec} \ , 
 \end{align}
 where in the above, relations with an $``="$ sign are exact relations, independent from the limit $k /q \rightarrow 0$, and relations with a $``\rightarrow"$ sign are only valid in the limit $k/q \rightarrow 0$.  We are concerned with the UV terms that we know must cancel, i.e. the ones proportional to $k^0 / q^0$.   so we define
 \begin{align} \label{f1uv}
  F^{(13)}_{1 , \text{UV}} ( \kvec , \qvec) &=  F^{(13)}_{2, \text{UV}} ( \kvec , \qvec) =  - F^{(13)}_{3, \text{UV}} ( \kvec , \qvec) =  - F^{(13)}_{4, \text{UV}} ( \kvec , \qvec)  =  - 2 \mu^2   P^{\rm in}_{\kvec} P^{\rm in}_{\qvec} \ , \\
 F^{(13)}_{5 , \text{UV}} ( \kvec , \qvec) &=  F^{(13)}_{6, \text{UV}} ( \kvec , \qvec) = 0  \ , \label{f3uv} 
 \end{align}
so that $F^{(13)}_{i,\text{UV}} ( \kvec, \qvec)= \lim_{k/q \rightarrow 0} F^{(13)}_{i} ( \kvec, \qvec)$ to order $k^0 / q^0$.  As we discussed above, one could also include the terms proportional to $k^2 / q^2$, which are degenerate with the counterterms, in the above definitions.  This would mean that the loop could be reliably computed using less precision because the overall size of the integrals is smaller.

The terms in $P_{22}$ are
 \begin{align} \label{firstf22uv}
 F^{(22)}_1 ( \kvec , \qvec ) & \rightarrow  \half \left(\frac{k^4}{q^4} ( 1 - 2 \mu^2)^2    + \mathcal{O} \left( \frac{k^5}{q^5} \right) \right) P^{\rm in}_{\qvec} P^{\rm in}_{\qvec} \ , \\
  F^{(22)}_2 ( \kvec , \qvec ) & \rightarrow  - \half \left(   \frac{k^4}{q^4} ( 1 - 2 \mu^2)  + \mathcal{O} \left( \frac{k^5}{q^5} \right) \right) P^{\rm in}_{\qvec} P^{\rm in}_{\qvec}  \ , \\
  F^{(22)}_3 ( \kvec , \qvec ) & \rightarrow  - \half \left(    \frac{k^4}{q^4}  ( 1 - 2 \mu^2) + \mathcal{O} \left( \frac{k^5}{q^5} \right) \right) P^{\rm in}_{\qvec} P^{\rm in}_{\qvec} \ , \\
    F^{(22)}_4 ( \kvec , \qvec ) & \rightarrow  \frac{1}{2}  \left(  \frac{k^4}{q^4}     + \mathcal{O} \left( \frac{k^5}{q^5} \right)\right) P^{\rm in}_{\qvec} P^{\rm in}_{\qvec} \ . \label{lastf22uv}
 \end{align}
All of these terms start proportional to $k^4 / q^4$, and so are degenerate with the stochastic counterterm.  As we mentioned above, we will not explicitly remove these terms from the integrand, although it is straightforward to do so.  These terms are typically small for a one-loop computation, so we do not expect them to have a large contribution anyway.  The terms proportional to $k^5/q^5$ are also proportional to an odd power of $\mu$, so we can eliminate them at the level of the integrand by using a quantity that is manifestly even for $\qvec \rightarrow - \qvec$.  With this in mind, we define the UV terms as
\be
 F^{(22)}_{1,{\rm UV}} ( \kvec , \qvec )  =  F^{(22)}_{2,{\rm UV}} ( \kvec , \qvec )  =  F^{(22)}_{3,{\rm UV}} ( \kvec , \qvec )  =  F^{(22)}_{4,{\rm UV}} ( \kvec , \qvec ) = 0 \ ,
 \ee
so that $F^{(22)}_{i,\text{UV}} ( \kvec, \qvec)= \lim_{k/q \rightarrow 0} F^{(22)}_{i} ( \kvec, \qvec)$ to order $k^3 / q^3$.

Before defining the UV-safe integrand, let us explicitly verify that the spurious UV terms do indeed cancel.  Since the structure of the would-be-needed counterterms is different, the spurious divergences in each diagram must cancel separately.\footnote{This follows from the fact that the spurious divergences with the same dependence on $k$ must cancel separately.}  Thus, in general, we must check that 
\begin{align} \label{anotherzero}
 \sum_{i=1}^6 T^{(13)}_i ( a , a_1 , a_2)  F^{(13)}_{i , \text{UV}}(\kvec , \qvec) =0 \  , 
 \end{align}
 and
 \begin{align} \label{22zero}
   \sum_{i=1}^4 T^{(22)}_i ( a , a_1 , a_2)  F^{(22)}_{i , \text{UV}}(\kvec , \qvec) = 0 \ ,
\end{align}
separately.  Because $F^{(22)}_{i,\text{UV}} = 0$, \eqn{22zero} is satisfied automatically.  For the $(13)$ terms, we must evaluate
\begin{align} \label{andanotherone}
 \sum_{i=1}^6 T^{(13)}_i ( a , a_1 , a_2)  F^{(13)}_{i , \text{UV}}(\kvec , \qvec) & = 2 \mu^2 K(a, a_1, a_2) \bar G^\delta_1( a , a_2) \Big(  - \bar G^\delta_1 ( a_2 , a_1) - \bar G^\delta_2 ( a_2 , a_1)  \\ 
 & \hspace{2in} + \bar G^\Theta_1 ( a_2 , a_1) + \bar G^\Theta_2 ( a_2 , a_1)    \Big) \nonumber
\end{align}
which is equal to zero as one can verify with the explicit expressions for the Green's functions \eqn{gdelta} - \eqn{gtheta}.  Now, we add and subtract \eqn{anotherzero} from $p_{1-\text{loop}}$ in a completely analogous way to last section.  Here we will skip the details and present the final expression, which is the same as the expression for $P_{1-\text{loop}}^{\text{IR-safe}}$ in \eqn{P1loop2}, except with the replacement
\be \label{firuv}
  F^{(13)}_{i,\text{IR-safe}} (\kvec , \qvec)   \rightarrow F^{(13)}_{i,\text{IR\&UV-safe}} (\kvec , \qvec)  \equiv F^{(13)}_i(\kvec , \qvec) -  F^{(13)}_{i , \text{IR}}(\kvec , \qvec) \, \Theta_{\rm H} ( k - q) - F^{(13)}_{i , \text{UV}}(\kvec , \qvec) \, \Theta_{\rm H} (  q - k ) \ ,
 \ee
 where we have multiplied $F^{(13)}_{i , \text{UV}}$ by the step function $\Theta_{\rm H} (  q - k )$ so that this term does not change the IR.  With this replacement, the fact that \eqn{andanotherone} is equal to zero ensures that we have simply added and subtracted zero from the integrand $p_{1-\text{loop}}$, so that the final integral is not changed.  However, the new integrand has the advantage that both the UV and IR parts which must cancel in the final expression for the one-loop power spectrum, are absent at the level of the integrand.  Explicitly, the formulae for the IR\&UV-safe integrand are
\begin{align}
& P^{\text{IR\&UV-safe}}_{ 1-\text{loop}}  ( a , k ) =  \int \momspmeas{q} \int_0^{a}  d a_2 \int_0^{a_2} d a_1 \Big(  p_{13}^{\text{IR\&UV-safe}} ( a , a_1 , a_2; \kvec ,  \qvec) \nonumber \\
& \hspace{3.5in} +   p_{22}^{\text{IR\&UV-safe}}(a, a_1 , a_2 ; \kvec , \qvec)  \Big) \ ,  \label{P1loop2uv}
\end{align}  
where
\begin{align} \label{deltap13uv}
 p_{13}^{\text{IR\&UV-safe}} ( a , a_1 , a_2; \kvec ,  \qvec)  &= \sum_{i=1}^6 T^{(13)}_i ( a , a_1 , a_2) \,  F^{(13)}_{i,\text{IR\&UV-safe}}(\kvec , \qvec) \ , \\
 p_{22}^{\text{IR\&UV-safe}} ( a , a_1 , a_2; \kvec ,  \qvec) & =   p_{22}^{\text{IR-safe}} ( a , a_1 , a_2; \kvec ,  \qvec)  \ , \label{deltap22uv}
\end{align}
 where $F^{(13)}_{i,\text{IR\&UV-safe}}(\kvec , \qvec)$ is given in~\eqn{firuv}.
 
 Now that we have completed our construction of the IR\&UV-safe integrand at one loop, let us briefly comment on how the procedure would generalize to higher loops.  As explained before, it is quite possible that it will be sufficient for comparison to data to evaluate just the one-loop terms with exact time dependence; higher loop contributions, which are smaller than the one-loop contributions, can then be computed with the EdS approximation without losing a relevant amount of precision on the overall result.  However, it is conceivable that one would like to check the difference between exact time dependence and the EdS approximation at two loops, or, for example, that one would like to compute the two-loop power spectrum in clustering quintessence for which there is no analogue of the EdS approximation.  The exact time dependence two-loop computation will be complicated by two main factors: there are more nested time integrals with various time orderings of Green's functions (this creates \emph{many} more independent diagrams), and there is an additional internal momenta which creates a more complicated region in the integration variables where IR and UV divergences occur.\footnote{Let us highlight a possible procedure. Regarding the Green's functions, one should first write all of the time integrals in a way that all of the diagrams have the same limits of integration of the time variables, as in \eqn{p22foryou} and \eqn{p13foryou}: this allows one to concentrate on the momentum dependent functions.  Then, as we commented in Footnote \ref{footnotef}, one could subtract all of the IR and UV divergences individually from the momentum dependent pieces (i.e. without mapping all of the divergences to the same point).  Because the divergences are subtracted from different parts of the integration region, it will be more difficult to explicitly check that the divergences cancel, but of course, we know that they must because of diffeomorphism invariance (for the IR contributions) or momentum conservation (for the UV contributions).  In any case, one could still do the explicit check.  A similar strategy will also help in tackling this problem for computations of the bispectrum or higher-point functions with exact time dependence.  We leave a full exploration of this topic to future work.}
 
 \section{Results} \label{resultssec}

\begin{figure}[htb!] 
\begin{center}
\includegraphics[width=13cm]{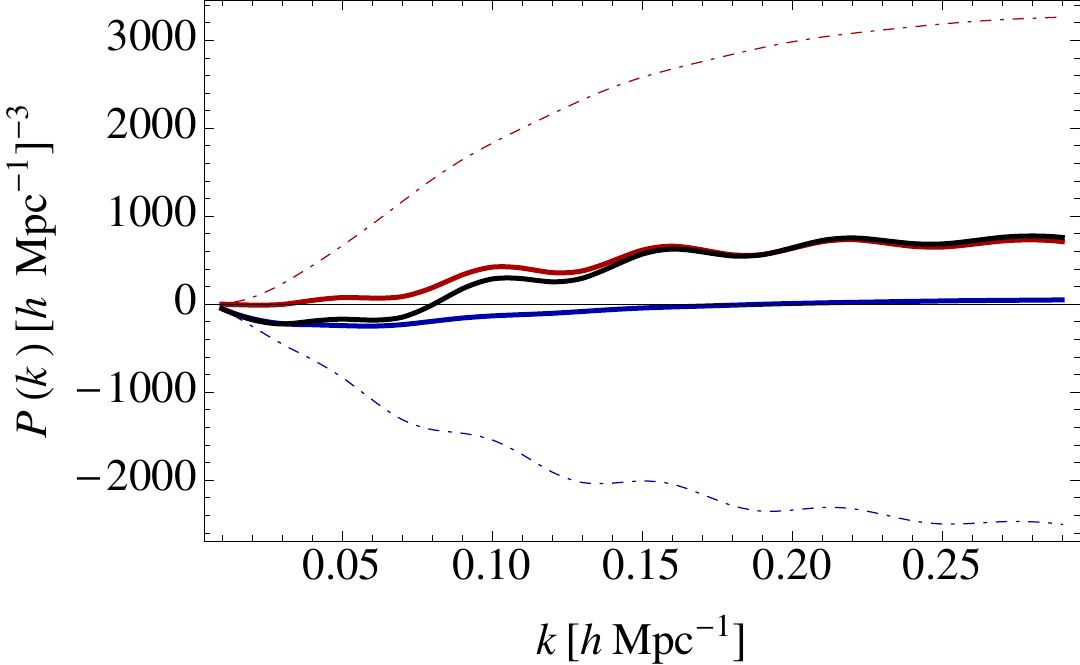}
\caption{Here, we compare the various contributions to $P_{1-\text{loop}}$ in clustering quintessence, both with and without the IR\&UV-safe integrand.  The solid curves use the IR\&UV-safe integrand, and the dot-dashed curves use the standard $p_{13}$ and $p_{22}$ integrands without any modifications.  The blue curves are the $(13)$ power spectra, the red curves are the $(22)$ power spectra, and the black curves are the total one-loop contributions.  One can see that for the individual contributions, $| P_{13}^{  \text{ IR\&UV-safe}}  | \ll  | P_{13}  |    $ and $| P_{22}^{  \text{ IR\&UV-safe}} |  \ll |  P_{22} |  $, but that the total contributions are essentially the same (they are indistinguishable in this plot, both contained in the black curve, see Figure~\ref{p1loopcompare2} for details).  Thus, one can compute the IR\&UV-safe integrals with much less numerical precision than the corresponding non-IR\&UV-safe integrals.   }   \label{p1loopcompare}
\end{center}
\end{figure}

In this section, we present the results of our computation, which we do for clustering quintessence at $z=0$  (described in more detail in \cite{Lewandowski:2016yce}) and use the cosmological parameters $\Omega_{m,0} = 0.27$, $\Omega_{D,0} = 0.73$ , $H_0 = 71 \, \text{km/s/Mpc}$, $\Delta_\zeta^2 = 2.42 \times 10^{-9}$, $n_s = 0.963$, and $w = -0.9$.  In order to implement the numerical computation, we compute each of the following terms separately
\begin{align}
& \int \momspmeas{q} \,  F^{(13)}_{i,\text{IR\&UV-safe} }(\kvec , \qvec) \\
& \int \momspmeas{q}  \, \Big(  F^{(22)}_{i,\text{IR\&UV-safe} }(\kvec , \qvec) +   F^{(22)}_{i,\text{IR\&UV-safe} }(\kvec , - \qvec) \Big) \\
&  \int_0^{a}  d a_2 \int_0^{a_2} d a_1 \, T^{(\sigma)}_i ( a , a_1 , a_2)  \ , 
\end{align}
multiply them together, and add the results for each $\kvec$.  The principle advantage of our approach is that each of the integrands of the momentum integrals is manifestly IR\&UV-safe, with both the $k^2 / q^2 $ and the $k / q$ divergences being canceled in the IR, and the $k^0 P_{11}(k)$ term canceled in the UV.  This means that at every step of the computation, we are adding numbers which are of the order of the final result, rather than relying on large numerical cancellations between terms.\footnote{As we mentioned, subleading UV-divergences can be removed in a similar way to what we do here, as already implemented in~\cite{Foreman:2015lca}.}  To see this, in Figure~\ref{p1loopcompare} we compare the following quantities
\begin{align}
P_{13}^{ \text{IR\&UV-safe}}  ( a , k ) &=  \int \momspmeas{q} \int_0^{a}  d a_2 \int_0^{a_2} d a_1 \,  p_{13}^{  \text{IR\&UV-safe} } ( a , a_1 , a_2; \kvec ,  \qvec) \ , \\
P_{22}^{ \text{IR\&UV-safe}}  ( a , k ) &=  \int \momspmeas{q} \int_0^{a}  d a_2 \int_0^{a_2} d a_1\,   p_{22}^{  \text{IR\&UV-safe}} (a, a_1 , a_2 ; \kvec , \qvec) \ ,  \\
 P_{13} ( a , k) & = \int \momspmeas{q} \int_0^{a}  d a_2 \int_0^{a_2} d a_1 \, \,  p_{13}(a, a_1 , a_2 ; \kvec , \qvec) \ , \\
P_{22} ( a , k) & = \int \momspmeas{q} \int_0^{a}  d a_2 \int_0^{a_2} d a_1 \, \,  p_{22}(a, a_1 , a_2 ; \kvec , \qvec) \ .
\end{align} 
From Figure~\ref{p1loopcompare}, it is clear that computing $P_{1-\text{loop}}$ without the IR\&UV-safe integrand involves a large cancellation: $| P_{22} | \approx | P_{13} | \gg | P_{1-\text{loop}} |$.  However, when using the IR\&UV-safe integrand, no cancellation is involved in the computation.  The goal of using IR\&UV-safe integrands is to have  $| P_{22}^{ \text{IR\&UV-safe}} | \approx | P_{13}^{ \text{IR\&UV-safe}} | \approx | P_{1-\text{loop}} |$ so that the same numerical precision that is desired in the end can be used to compute the separate terms.  In fact, the situation here is even better: for low $k$, $ | P_{13}^{ \text{IR\&UV-safe}} | \approx  | P_{1-\text{loop}} |$ and $ | P_{22}^{ \text{IR\&UV-safe}} | \ll  | P_{1-\text{loop}} |$, while for higher $k$ the contributions switch so that $ | P_{22}^{ \text{IR\&UV-safe}} | \approx  | P_{1-\text{loop}} |$ and $ | P_{13}^{ \text{IR\&UV-safe}} | \ll  | P_{1-\text{loop}} |$.  We remind the reader that if an individual contribution is much \emph{less} than the final result, then one can in principle use even less numerical precision to compute that contribution.

\begin{figure}[htb!] 
\begin{center}
\includegraphics[width=16cm]{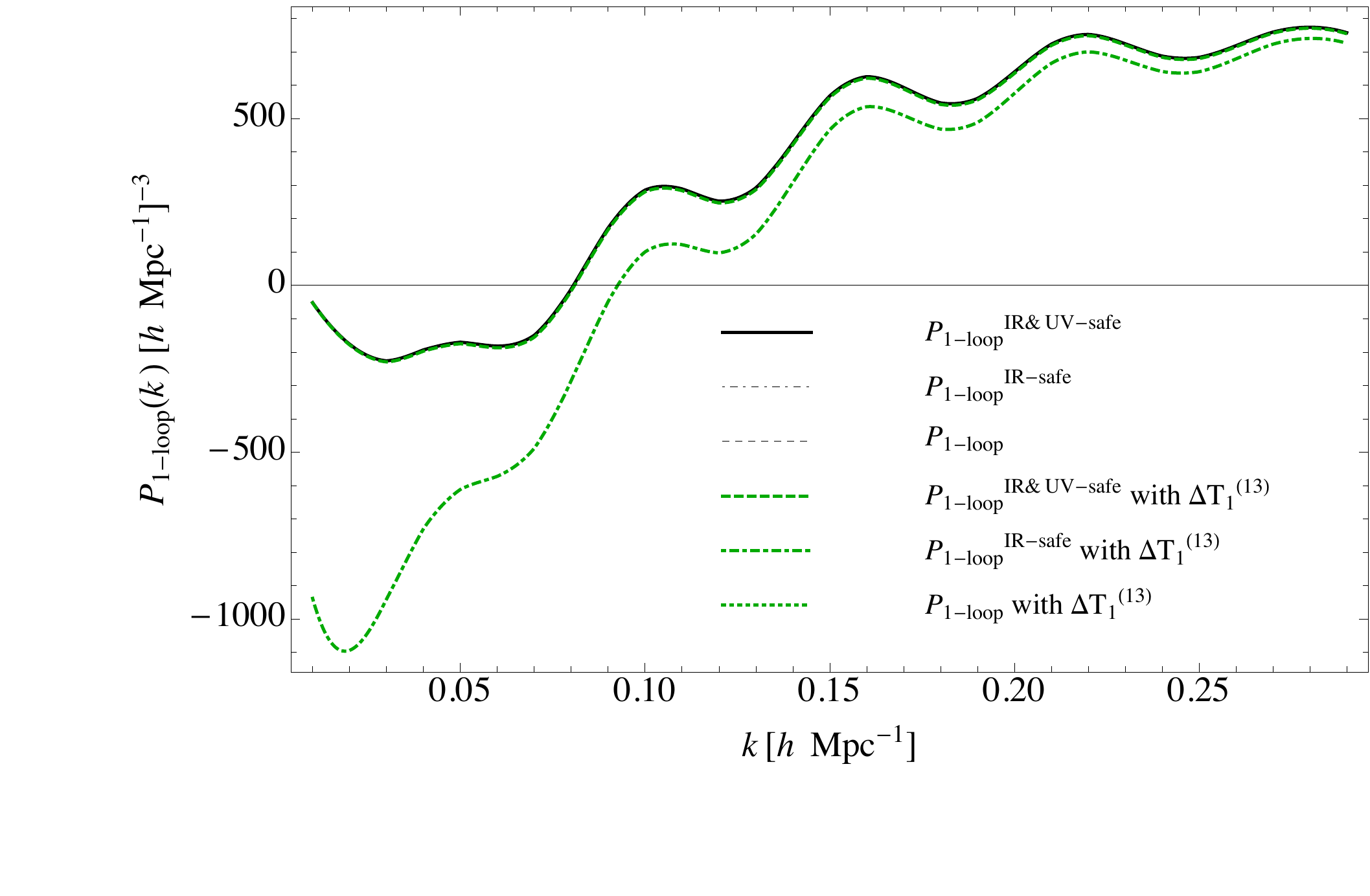} \hspace{1in}\\
\caption{In this figure, we show the effect of imprecisely computing the integrals of the time-dependent factors $T^{(\sigma)}_i$.  In particular, for illustration, we have changed one of the time dependent coefficients, $T^{(13)}_1$, by 1\%.  This plot contains six curves.  The black solid curve is $P^{\text{IR\&UV-safe}}_{ 1-\text{loop}}$, the black dot-dashed curve is $P^{\text{IR-safe}}_{ 1-\text{loop}}$, the black dashed curve is $P_{ 1-\text{loop}}$. Since these curves appear as the single solid curve (they are indistinguishable in this plot), we deduce that the numerical integration of the time coefficients is done sufficiently.  All of the green curves have $T^{(13)}_1$ changed by 1\%: the dashed curve is $P^{\text{IR\&UV-safe}}_{ 1-\text{loop}}$, the dot-dashed is $P^{\text{IR-safe}}_{ 1-\text{loop}}$, and the dotted is $P_{ 1-\text{loop}}$ (they are indicated in the legend as ``with $\Delta T_1^{(13)}$'').  We can see that of the curves with an incorrect $T^{(13)}_1$, $P^{\text{IR-safe}}_{ 1-\text{loop}}$ and $P_{ 1-\text{loop}}$ (which both appear as the dot-dashed curve because they are overlaid) are greatly affected at low $k$, while $P^{\text{IR\&UV-safe}}_{ 1-\text{loop}}$ is essentially unchanged (see Figure~\ref{p1loopcompare22} for more details).  This shows that in this computation, using the UV-safe integrand is the most important, although we expect spurious IR effects to be more of a nuisance in a two-loop or higher order computation.          }  \label{p1loopcompare2}
\end{center}
\end{figure}

\begin{figure}[htb!] 
\begin{center}
\includegraphics[width=7.5cm]{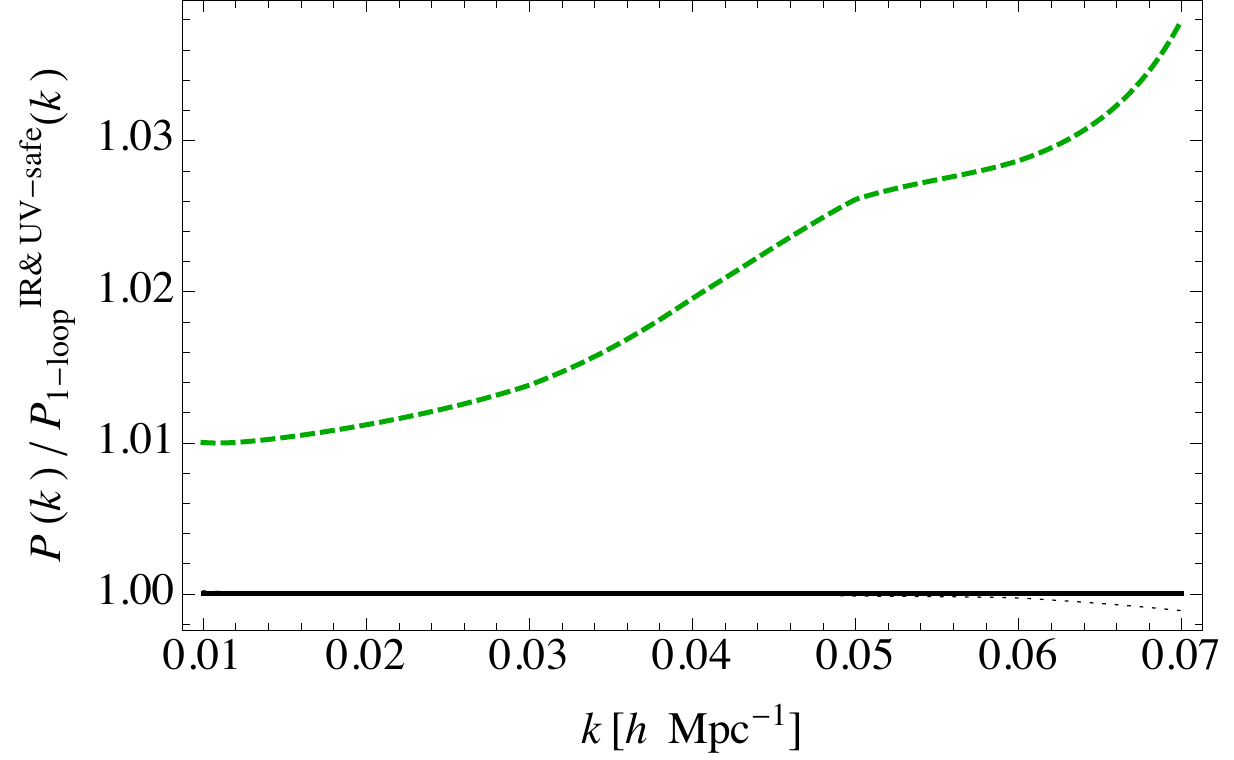} \hspace{.1in} \includegraphics[width=7.7cm]{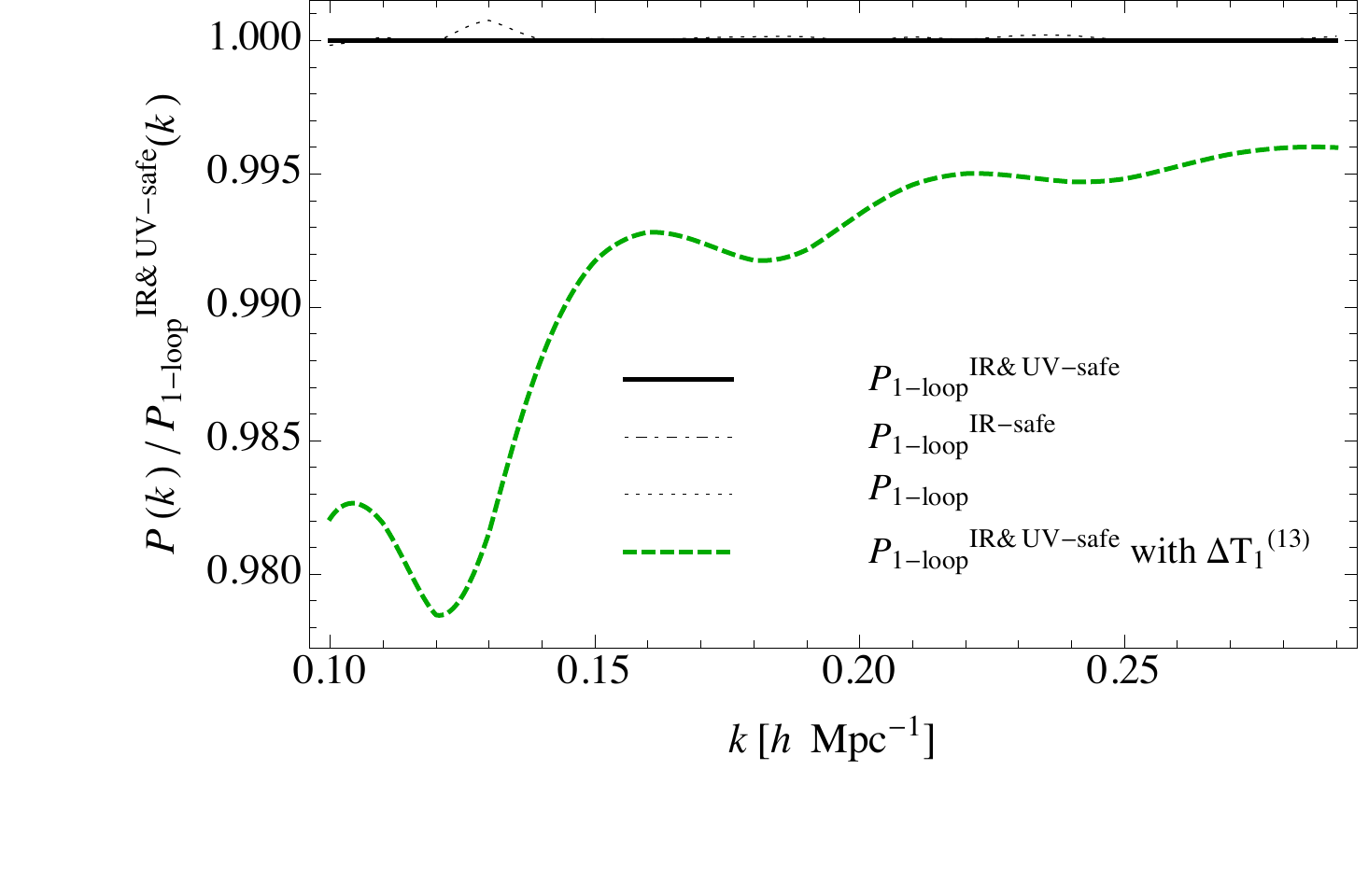}  \\
\vspace{.1in}
\includegraphics[width=7.5cm]{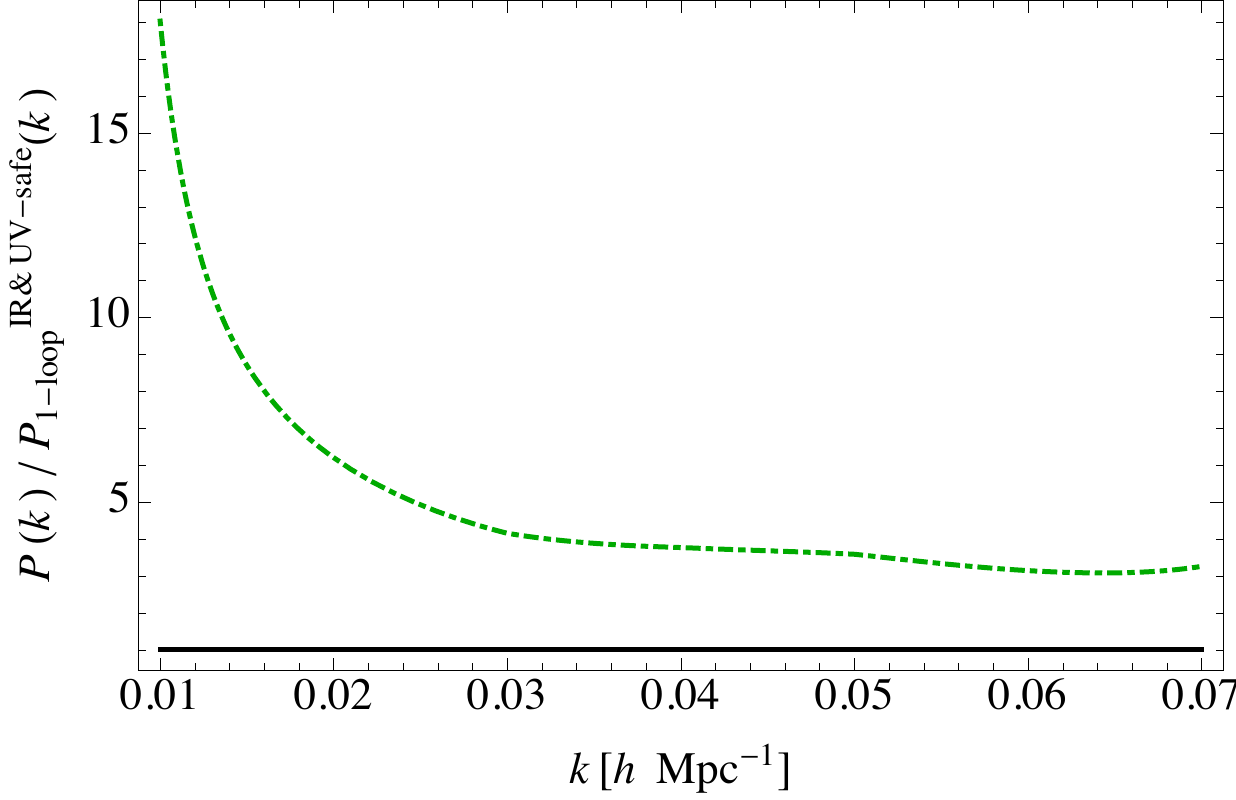} \hspace{.1in} \includegraphics[width=7.7cm]{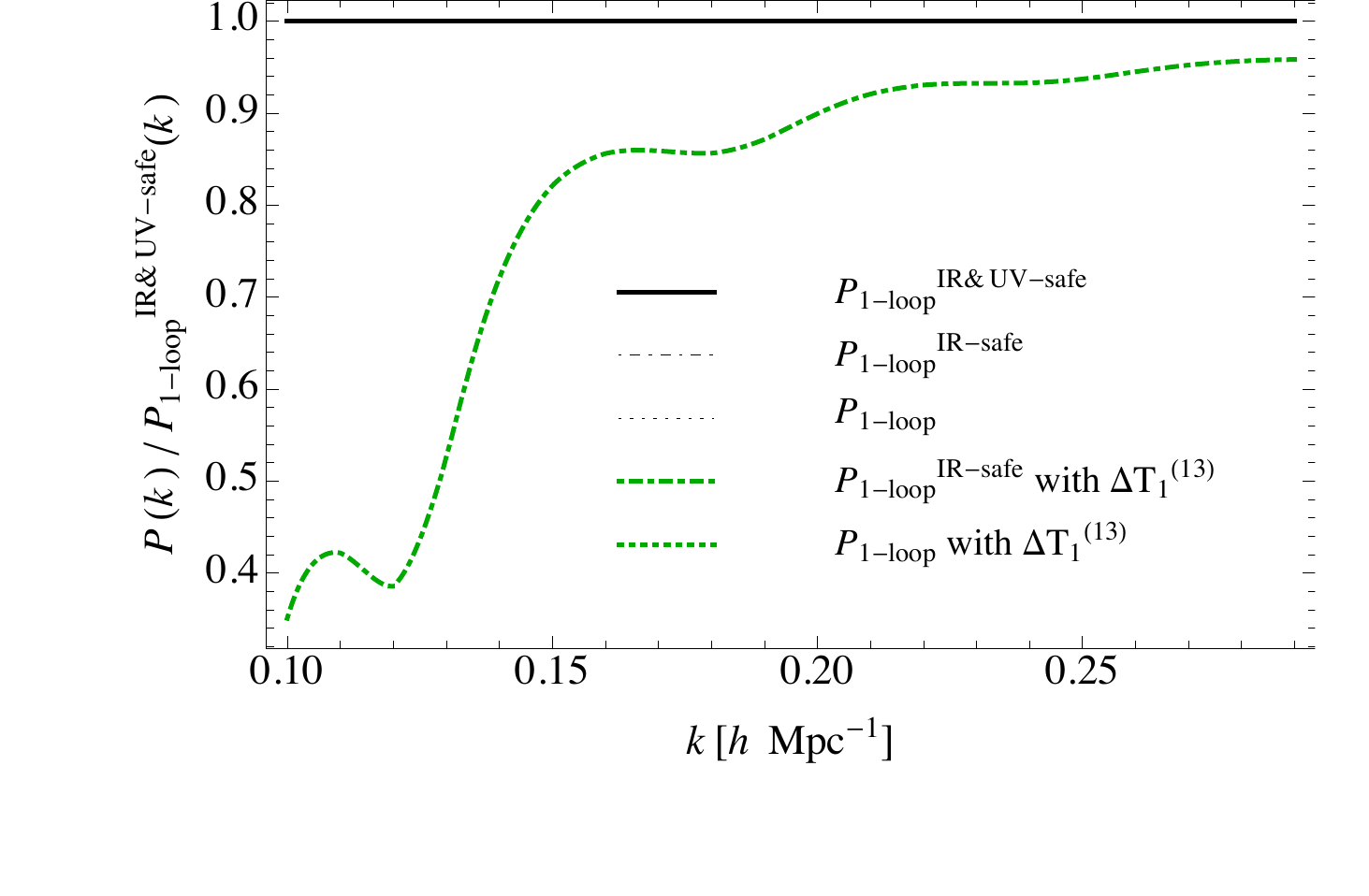}  
\caption{This figure shows the same information as Figure~\ref{p1loopcompare2}, but in more detail, and the labeling of the curves is exactly the same as in Figure~\ref{p1loopcompare2} (legends are provided on the right-hand side of each row).  In this plot, we divide the various methods of computation by $P^{\text{IR\&UV-safe}}_{ 1-\text{loop}}$ (with the numerical coefficients computed correctly) to study in more detail the size of the various effects (this also explains why we do not plot between $k = 0.07 \unitsk$ and $k = 0.1 \unitsk$, where $P^{\text{IR\&UV-safe}}_{ 1-\text{loop}} \rightarrow 0$ and causes the curves to blow up).  Computations for which we have changed $T^{(13)}_1$ by 1\% are indicated in the legend as ``with $\Delta T_1^{(13)}$.'' From the top two plots, we see that the effect of changing $T^{(13)}_1$ by 1\% in $P^{\text{IR\&UV-safe}}_{ 1-\text{loop}}$ (the green dashed curve in the top plots) is a few percent.  However, in the lower two plots, we see that the effect of changing $T^{(13)}_1$ by 1\% in  $P^{\text{IR-safe}}_{ 1-\text{loop}}$ and  $P_{ 1-\text{loop}}$ (respectively the green dot-dashed and green dotted curves in the lower plots) is much more dramatic: larger than a factor of 5 for low $k$ and between 10\% and 70\% at higher $k$.  This shows that in this computation, using the UV-safe integrand is most important, although we expect spurious IR effects to be more of a nuisance in a two-loop or higher order computation.    }  \label{p1loopcompare22}
\end{center}
\end{figure}

\begin{figure}[htb!] 
\begin{center}
\includegraphics[width=7.5cm]{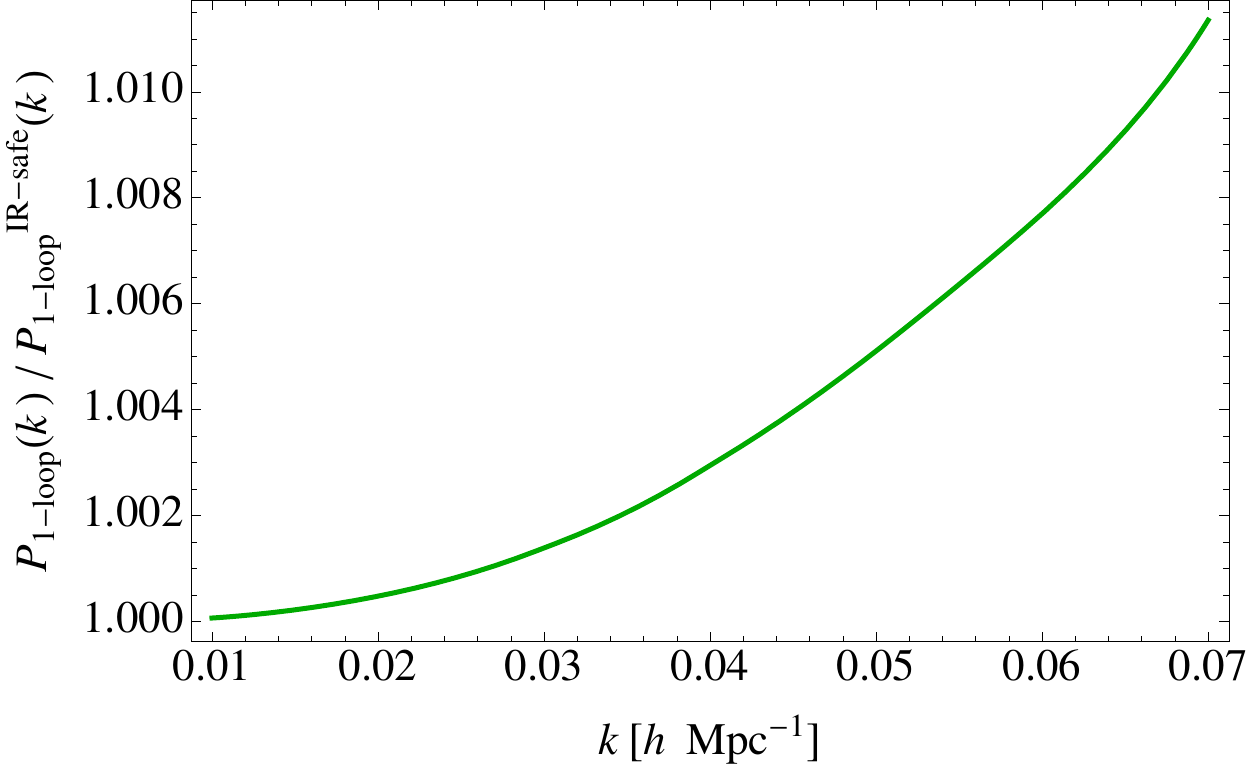} \hspace{.1in} \includegraphics[width=7.7cm]{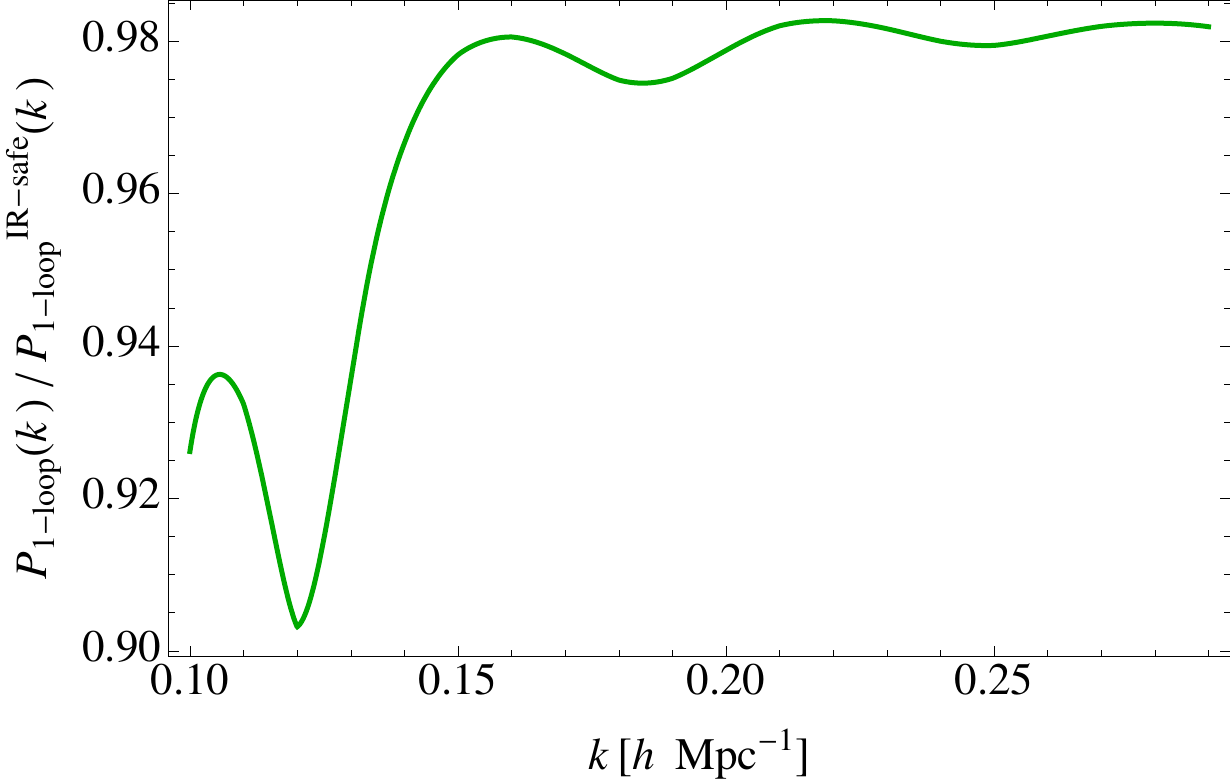}  
\caption{The difference between using $P_{ 1-\text{loop}}$ and $P^{\text{IR-safe}}_{ 1-\text{loop}}$, where $T^{(13)}_1$ has been changed by 1\% in each expression, is about 1\% at low $k$, and is between 2\% and 7\% at higher $k$, which is still non-negligibly boosted from the expected one percent.  As can be seen in Figure~\ref{p1loopcompare22}, both $P_{ 1-\text{loop}}$ and $P^{\text{IR-safe}}_{ 1-\text{loop}}$ (because they do not use UV-safety) are very different from the more precise answer $P^{\text{IR\&UV-safe}}_{ 1-\text{loop}}$ (by more than a factor of $5$ at low $k$, and between 10\% and 70\% at higher $k$).  Figure~\ref{lastplotlabel} isolates the effect of IR-safety, and shows that it is between a 2\% and 7\% effect.         }  \label{lastplotlabel}
\end{center}
\end{figure}

Then, in Figures~\ref{p1loopcompare2}, \ref{p1loopcompare22} and \ref{lastplotlabel}, we compare the various ways of computing $\Poneloop$: using the standard $\Poneloop$ without any IR or UV subtractions, using $P^{\text{IR-safe}}_{ 1-\text{loop}} $, and using $P^{\text{IR\&UV-safe}}_{ 1-\text{loop}}$.  In order to see how sensitive these three methods of computation are to the precise evaluation of the time dependent coefficients, we also plot the above three computations after changing the value of one of the time dependent coefficients, $T^{(13)}_1$, by 1\%.  In this one-loop computation, we find that overall it is more important to use the UV-safe integrand than the IR-safe integrand: when changing $T^{(13)}_1$ by 1\%, the curves which are not UV-safe are wrong by more than a factor of $5$ at low $k$, and are wrong by between 10\% and 70\% at higher $k$, while the curve which uses IR\&UV-safety is wrong by a few percent both at low $k$ and high $k$ (this is the expected change since we changed one of the terms by 1\%).\footnote{Of course, the sensitivity to using or not using UV-safety depends explicitly on the UV cutoff of the loop integrals, which we take to be $\Lambda_{\text{UV}} = 10 \unitsk$.  On the other hand, because of the natural IR cutoff due to the matter-radiation equality scale near $k_{\rm eq} \sim 0.01 \unitsk$, the sensitivity to using or not using IR-safety should be essentially independent of the IR cutoff $\Lambda_{\rm IR}$ used in the loop integrals, as long as $\Lambda_{\rm IR} \ll k_{\rm eq}$.}  In particular, Figure~\ref{lastplotlabel} shows that the difference between using only the IR-safe integrand $P^{\text{IR-safe}}_{ 1-\text{loop}} $ and using the unimproved $\Poneloop$ is between 2\% and 7\%, which is still non-negligibly boosted from the expected one percent.  As discussed above, this sensitivity arises because each contribution to $P_{1-\text{loop}} ( a , k) $, $P_{13}(a,k)$ and $P_{22}(a,k)$, are themselves a sum of terms which are products of a time integral and a momentum integral.  We showed that the cancellation of IR and UV divergences involves many of these terms, which generically have different time dependences, together.  Thus, only with very large precision of the time integrals are the IR and UV divergences guaranteed to cancel; even a small numerical error for the time dependent coefficients can produce a large overall error, simply because the individual IR and UV contributions are large.  None of this is an issue if the IR\&UV-safe integrands are used.  Although the effect of not using IR-safety is sizable but not very large in the one-loop computation that we present here, it is expected that the spurious IR parts of the loop integrals will be much more of a nuisance in a two-loop or higher order calculation, where the time- and momentum- integrals are more complex. A similar consideration applies to the UV divergencies, as loops become more divergent at higher order.

\begin{figure}[htb!] 
\begin{center}
\includegraphics[width=8.4cm]{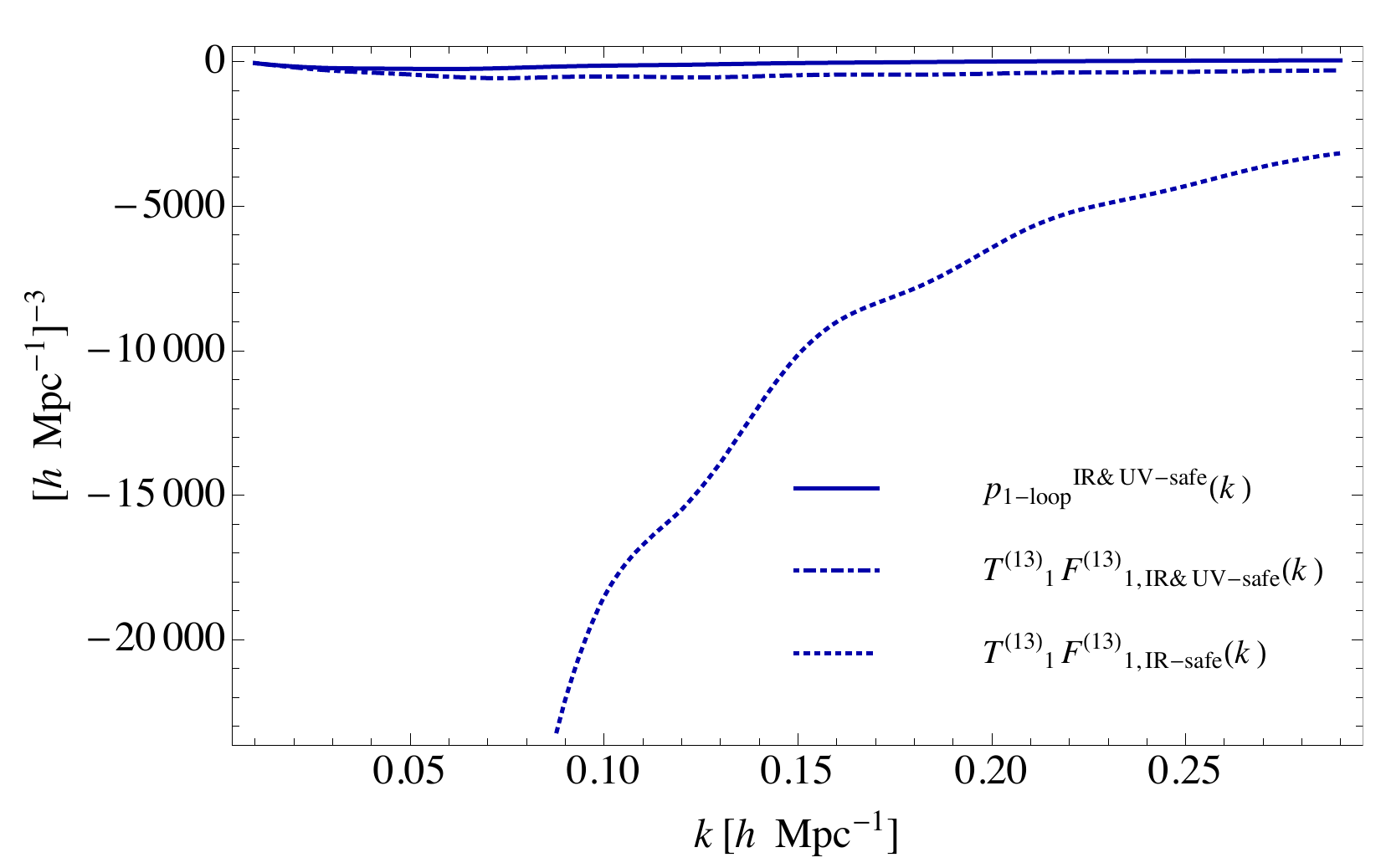}  \includegraphics[width=7.8cm]{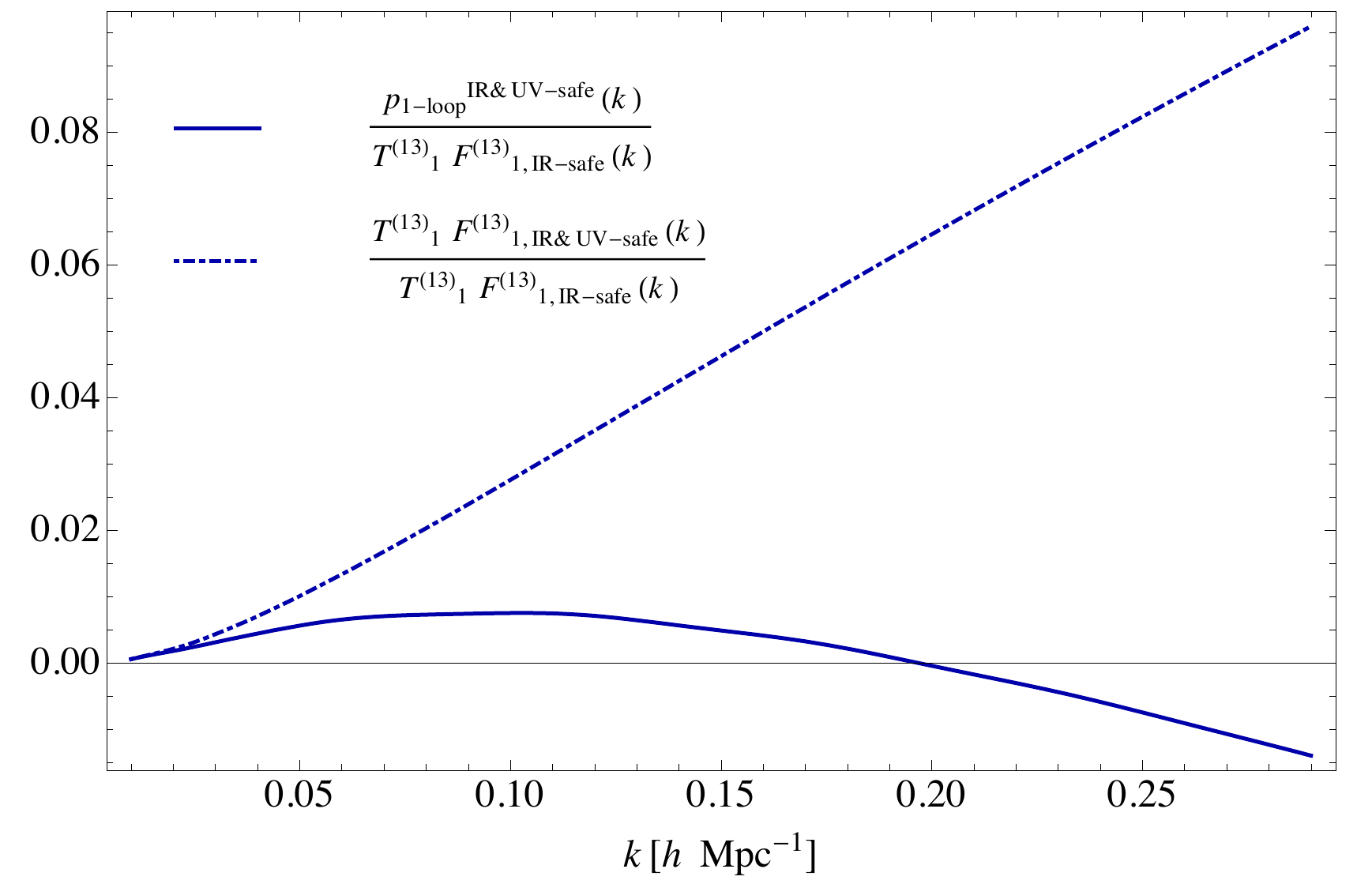}  
\caption{In this figure, we show the the size of the UV terms that we removed from $P_{13}$ by looking at the contribution from $T^{(13)}_1 F^{(13)}_1$ in particular.  It is clear that the term without the UV-safe integrand is much larger than the one with the UV-safe integrand.  Without using the UV-safe integrand, a very large part of $T^{(13)}_1( a , a_1 , a_2) F^{(13)}_{1,\text{IR-safe}   } ( \kvec , \qvec)$ must be canceled by another term $T^{(13)}_i( a , a_1 , a_2)F^{(13)}_{i,\text{IR-safe}   }( \kvec , \qvec) $ for $i\neq1$ in order to get down to the final answer of $p_{1-\text{loop}}^{\text{IR\&UV-safe}}$; this cancellation has to happen at the level of $5\times 10^{-3}$ around $k = 0.05 \unitsk$.  Thus, if one does not use the UV-safe integrand, the final answer is much more sensitive to the precision with which the integrals of the time dependent coefficients $T^{(13)}_i$ are determined.}             \label{uvdivplot}
\end{center}
\end{figure}

Finally, in Figure~\ref{uvdivplot} we explicitly show the large cancellation between terms in $P_{13}$ that must happen in the UV by looking at the contribution $T^{(13)}_1 F^{(13)}_1$ in particular.  We see that the IR\&UV-safe integrand is much smaller than the integrand with only IR-safety ($F^{(13)}_{1,\text{IR\&UV-safe}} / F^{(13)}_{1,\text{IR-safe}} \approx 0.005$ at $k = 0.05 \unitsk$), and that $T^{(13)}_1 F^{(13)}_{1,\text{IR\&UV-safe}}$ is much closer to the final answer $p_{1-\text{loop}}^{\text{ IR\&UV-safe}}$ than the integrand without UV safety.  This means that without using the UV-safe integrands, individual terms in $P_{13}$ must cancel to the level $5 \times 10^{-3}$, and so the final answer is very sensitive to the precision with which the time dependent coefficients $T^{(13)}_i$ are determined.  This means that we can greatly speed up numerical computation time by using the UV\&IR-safe integrands (which are expected to help even more in a two-loop or higher order computation), where these problems are not present.

%
%
%

\section{ Precision comparison} \label{preccompsec}
\begin{figure}[htb!] 
\begin{center}
\includegraphics[width=17cm]{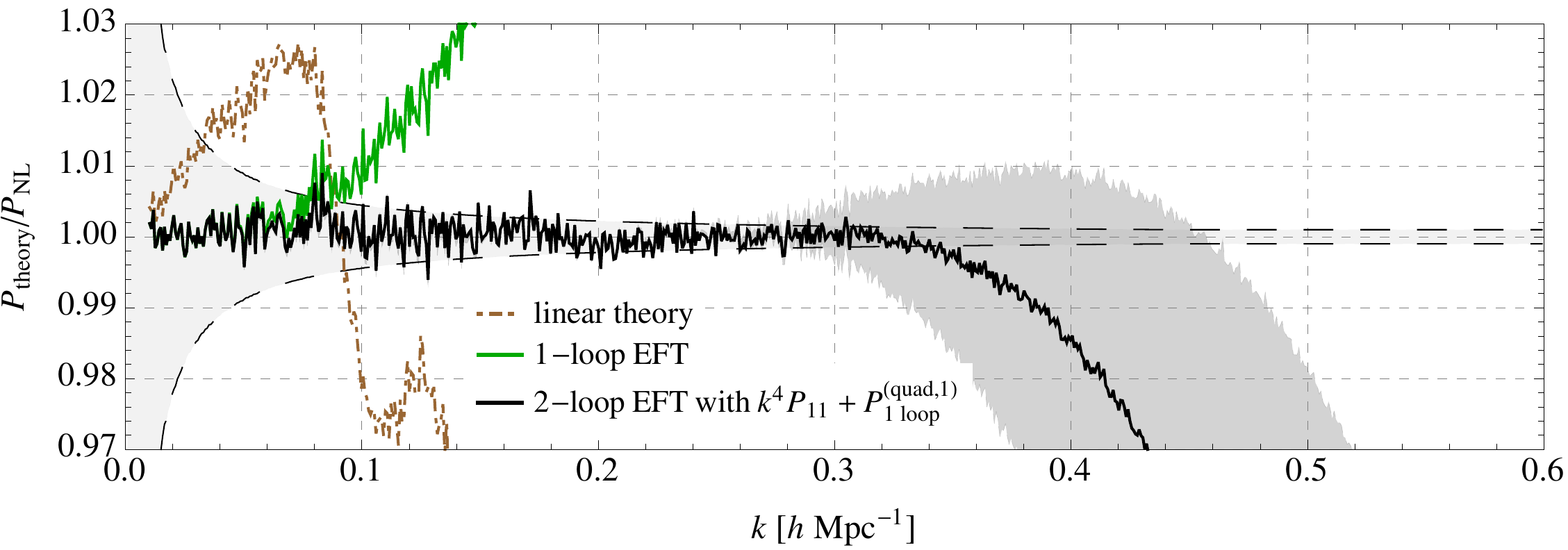} 
\caption{ We present the precision comparison of the two-loop dark-matter power spectrum to the Dark Sky $N$-body simulation.  As an improvement to the computation done in~\cite{Foreman:2015lca}, we use $P^{\text{IR\&UV-safe}}_{ 1-\text{loop}}$ with exact time dependence instead of the $P_{\text{EFT-1-loop}}$ with the EdS approximation, which was used in~\cite{Foreman:2015lca}.  That is, all one-loop terms are computed using the exact time dependence presented in this paper, but two-loop terms are computed using the EdS approximation.  The results are very similar to the ones obtained in~\cite{Foreman:2015lca}, but slightly better.      }           \label{preccomp}
\end{center}
\end{figure}

In this section, we compare the two-loop power spectrum for dark matter in the EFTofLSS at $z=0$ to the Dark Sky $N$-body simulation~\cite{Skillman:2014qca}, in a $\Lambda$CDM cosmology with cosmological parameters $\Omega_{m,0} = 0.295$, $\Omega_{\rm baryon} = 0.0468$, $\Omega_{\Lambda,0} = 0.705$, $h = 0.688$, $n_s = 0.9676$, and $\sigma_8 = 0.835$.\footnote{\url{http://darksky.slac.stanford.edu/} }  This precision comparison was originally done in~\cite{Foreman:2015lca} using the EdS approximation for all time dependence in the power spectrum (i.e. for linear, one-loop, and two-loop terms).  In Figure~\ref{preccomp}, we provide the same computation done in \cite{Foreman:2015lca}, but we use the $P^{\text{IR\&UV-safe}}_{ 1-\text{loop}}$ with exact time dependence (which we computed above), instead of the $P_{\text{EFT-1-loop}}$ with the EdS approximation used in \cite{Foreman:2015lca}.  In other words, in Figure~\ref{preccomp}, all one-loop terms are computed using the exact time dependence presented in this paper, but two-loop terms are computed using the EdS approximation.  We refer the reader to~\cite{Foreman:2015lca} for all the details of this computation.  In Figure~\ref{preccomp}, we see that the result is very similar to the one obtained in~\cite{Foreman:2015lca}, but slightly better.  In Appendix~\ref{fittingdetails}, we give details about the determination of the coupling constants used in Figure~\ref{preccomp}.  The values of the counterterm parameters are given as
\begin{align}
c_{s(1)}^2 \simeq 0.57 \left( \frac{ \knl}{ 2 \unitsk} \right)^2 \ , \hspace{.1in}
c_{1} \simeq -0.97 \left( \frac{ \knl}{ 2 \unitsk} \right)^2 \ , \hspace{.1in}
c_{4} \simeq -6.6 \left( \frac{ \knl}{ 2 \unitsk} \right)^4 \ . 
\end{align}

%
%
%

\section{Conclusion} \label{conclusionsec}
Large-scale structure surveys may very well be the next leading sources of cosmological information.  Because most modes are concentrated on short scales, it is important to understand large-scale structure observables in the mildly non-linear regime.  In order to accomplish this, the EFTofLSS has been developed to systematically and controllably include the effects of gravitational clustering in the UV on the mildly non-linear regime of interest.  This approach increases our understanding in two ways: first, it extends the maximum $k$ at which we understand the theory, and second, for $k \lesssim \knl$, it allows us to compute observables to a very high precision by including more and more loops.  So far, most computations in the EFTofLSS have used the so-called EdS approximation to solve for the time dependence of the loop contributions.  Because this approximation was known to be accurate to less than~$1\%$, and because the aim of previous computations had been about $1\%$ accuracy, the EdS approximation was perfectly fine.  However, since the ultimate goal of the EFTofLSS is precision computation, it is conceivable that less than $1\%$ accuracy will be desirable in the future, in which case one will be forced to use the exact time dependence routinely (at least on the lower order loops).  As an alternative to using the exact time dependence in $\Lambda$CDM, one could also improve the EdS approximation by expanding the time dependence around EdS.  Because EdS is such a good approximation in $\Lambda$CDM, this should be a very quickly converging expansion, with an expansion parameter of $\mathcal{O}(1/100)$ (we highlighted a procedure to do this in Footnote~\ref{footnote:perturbative_exact_time}).  However, because the exact time dependence is not too complicated at one loop, and because we are also interested in clustering quintessence in this work (for which there is no analogue of the EdS approximation), we leave an exploration of this direction to future work.

There is a small technical challenge to using the exact time dependence, though.  Because the diagrams become more complicated, $P_{13} ( a , k)$ and $P_{22} ( a , k) $ are each a sum of terms which are products of a function of momenta, $F^{(\sigma)}_i ( \kvec , \qvec)$, and a function of times, $T^{(\sigma)}_i ( a, a_1, a_2)$.  Each contribution is then separately integrated over $d^3 q$ and $da_2 \, da_1$, and then the results are added together.  In each $F^{(\sigma)}_i$, there are both IR and UV divergent terms which must ultimately cancel in the full equal-time one-loop expression (the IR cancellation is due to the equivalence principle, the UV one to matter conservation), but inexact evaluation of the time integrals can spoil this cancellation.  Thus, one could be left with spurious numerical contributions to the integrand of $P_{1-\text{loop}}$.  Said another way, one has to compute the numerical time integrals (and of course also the momentum integrals, which have spurious divergencies in different regions of the integration over $d^3 q$) to a very high precision to make sure that the final result is not dominated by these spurious contributions as $q/k \rightarrow 0$ and as $k/q \rightarrow 0$.  This kind of problem would defeat the purpose of using the exact time dependence in the first place.

However, since we know that these IR and UV terms must cancel in the final result, we can re-write the standard integrands for the exact time-dependent diagrams $P_{13}$ and $P_{22}$ into a form in which these IR and UV divergent terms never enter the numerical computation at all: this is the IR\&UV-safe integrand given by \eqn{P1loop2uv}, \eqn{deltap13uv} and \eqn{deltap22uv}.  Contrary to the previously supplied IR-safe integrand for the EdS approximation \cite{Carrasco:2013sva}, for which a single integrand could be used for the one-loop computation, the non-trivial time dependence considered in this paper forces us to write each of the many \emph{contributions} to $P_{1-\text{loop}}$ in a manifestly IR\&UV-safe way.  We find that, in the one-loop computation presented here, UV-safety is more important than IR-safety, although we expect the effects of spurious IR terms to be much more of a nuisance for two-loop and higher order computations.

 While doing this, we have extended the results of this paper to the adiabatic mode in the dark matter plus clustering quintessence system by including the non-trivial time dependent factor in the continuity equation, given by a function $C(a)$ in the equations of motion (where $C(a) = 1$ for $\Lambda$CDM).  In that system, because it is in the limit of small speed of sound ($c_s^2 \rightarrow 0$) of quintessence, there is really only one mode, the adiabatic mode $\delta_A$.  Because the equivalence principle is not violated, the effects of the bulk velocity and the gradient of the gravitational potential can be removed by a diffeomorphism.  Thus, we recover that the full equal-time power spectrum is free of IR divergences, and thus also establish that the consistency conditions are satisfied.  
 
 Finally, we have presented an improved precision comparison of the two-loop dark-matter power spectrum to the Big Sky $N$-body simulation.  In order to make our computation more precise, we have used $P^{\text{IR\&UV-safe}}_{ 1-\text{loop}}$ with exact time dependence instead of $P_{\text{EFT-1-loop}}$ with the EdS approximation that was used in~\cite{Foreman:2015lca}.  We found that the results are very similar, although there is indeed an improvement in the right direction.

\section*{Acknowledgments}

We would like to thank Simon Foreman for initial collaboration and illuminating discussions related to this project, and for helping us to produce the plots of Section~\ref{preccompsec} and Appendix~\ref{fittingdetails}.  We would also like to thank Azadeh Maleknejad for initial collaboration on this project.  L.S. is partially supported by DOE Early Career Award DE-FG02-12ER41854.

\section*{Appendices}

\appendix

\section{Reference formulae} \label{refform}
In this appendix, we provide some relevant formulae from \cite{Lewandowski:2016yce} for reference.  To avoid clutter, we have removed the ``$A$" subscript from the adiabatic fields $\delta_A$ and $\Theta_A$, so that $\delta$ and $\Theta$ refer to the adiabatic fields.  The equations of motion for the adiabatic mode $\delta_{\kvec}$ and $\Theta_{\kvec}$ in clustering quintessence are ($\Lambda$CDM is obtained by setting $C(a)=1$)
\begin{align} \label{lineareqs}
&a\delta'_{\vk}-f_{+}\Theta_{\vk}=\frac{(2\pi)^{3}f_{+}}{C(a)}\iint \frac{d^3q_1}{(2\pi)^{3}}\frac{d^3q_2}{(2\pi)^{3}} \delta_{D}(\vk-\vq_1-\vq_2)\alpha(\vq_1,\vq_2)\Theta_{\vq_1}\delta_{\vq_2} \ ,\\
&a\Theta'_{\vk}-f_{+}\Theta_{\vk}-\frac{f_{-}}{f_{+}}(\Theta_{\vk}-\delta_{\vk})=\frac{(2\pi)^{3}f_{+}}{C(a)}\iint \frac{d^3q_1}{(2\pi)^{3}}\frac{d^3q_2}{(2\pi)^{3}}\delta_{D}(\vk-\vq_1-\vq_2)\beta(\vq_1,\vq_2)\Theta_{\vq_1}\Theta_{\vq_2}  \ ,  \label{lineareqs2}
\end{align}
 where $f_\pm ( a ) = a \, \partial_a D_\pm(a) / D_\pm(a)$, $\alpha( \qvec_1 , \qvec_2)$ and $\beta ( \qvec_1 , \qvec_2)$ are the standard dark-energy interaction vertices given in \eqn{alphadef2} and \eqn{betadef2}, and $D_+$ is the growing solution to the second-order linear system for $\delta_{\kvec}$ defined by \eqn{lineareqs} and \eqn{lineareqs2}.  The non-trivial time dependent factor in the clustering quintessence system is given by 
 \be \label{thisisc}
 C(a) = 1 + (1+w) \frac{\Omega_{D,0}}{\Omega_{m,0}} \left( \frac{a}{a_0} \right)^{-3w}   \ ,
 \ee
 where $\Omega_{D,0}$ is the quintessence energy-density fraction today, $\Omega_{m,0}$ is the dark-matter energy-density fraction today, and $w$ is the equation of state for dark energy.  We can solve this system with Green's functions by expanding $\delta = \delta^{(1)} + \delta^{(2)} + \delta^{(3)} + \delta^{(ct)} + \dots$ (where $\delta^{(ct)}$ is the one-loop counterterm contribution) and finding 
 \begin{align}  \label{dtGreen}
 &\delta^{(n)}_{\vk}=\int^a_0 d\ta \bigg(G^{\delta}_{1}(a,\ta)S^{(n)}_1(\ta,\vk)+G^{\delta}_{2}(a,\ta)S^{(n)}_2(\ta,\vk)\bigg) \ ,\\
  &\Theta^{(n)}_{\vk}=\int^a_0 d\ta \bigg(G^{\Theta}_{1}(a,\ta)S^{(n)}_1(\ta,\vk)+G^{\Theta}_{2}(a,\ta)S^{(n)}_2(\ta,\vk)\bigg) \ , \label{dtGreen2}
  \end{align}
where $G^\delta_1$, $G^\delta_2$, $G^\Theta_1$, and $G^\Theta_2$ are the Green's functions for the system ($G^\delta_1$ encodes the response of $\delta$ to a perturbation to the continuity equation, $G^\delta_2$ encodes the response of $\delta$ to a perturbation to the Euler equations, and similarly for $\Theta$), and the source terms $S^{(n)}_i$ are the $n$-th order expansion of the right-hand sides of \eqn{lineareqs} and \eqn{lineareqs2} and are given explicitly in~\cite{Lewandowski:2016yce}.  Using \eqn{dtGreen} and \eqn{dtGreen2} in \eqn{lineareqs} and \eqn{lineareqs2}, we find that the four Green's functions are specified by the following equations
 \begin{align}
 &a \frac{d G^{\delta}_{\sigma}(a,\ta)}{da}-f_{+}(a)G^{\Theta}_{\sigma}(a,\ta)=\lambda_{\sigma}\delta(a-\ta) \ ,  \label{Green} \\
 &a \frac{d G^{\Theta}_{\sigma}(a,\ta)}{da}-f_{+}G^{\Theta}_{\sigma}(a,\ta)-\frac{f_{-}(a)}{f_{+}(a)}\bigg(G^{\Theta}_{\sigma}(a,\ta)-G^{\delta}_{\sigma}(a,\ta)\bigg)=(1-\lambda_{\sigma})\delta(a-\ta) \ ,
 \end{align}
where $\lambda_\sigma$ is $$\lambda_1=1 \quad \textmd{and} \quad \lambda_2=0,$$ $\sigma=1,2$, and $\delta(a-\ta)$ is the Dirac delta function. The retarded Green's functions satisfy the boundary conditions 
\begin{align}  \label{bound}
& G^{\delta}_\sigma(a,\tilde a)  =  0 \quad \quad  \text{and}   \quad\quad G^{\Theta}_\sigma(a, \tilde a)=0  \quad \quad \text{for} \quad \quad \tilde a > a  \ , \\
 &G^\delta_\sigma ( \tilde a , \tilde a ) = \frac{\lambda_\sigma}{\tilde a}  \quad \hspace{.06in} \text{and} \hspace{.2in}  \quad G^{\Theta}_{\sigma} ( \tilde a  , \tilde a ) = \frac{(1 - \lambda_\sigma)}{\tilde a}  . \label{bound2}
\end{align}
We can then construct the Green's functions in the usual way using the linear solutions and the Heaviside step function, $\Theta_{\rm H} (a-\tilde a)$, and imposing the boundary conditions \eqn{bound} and \eqn{bound2}.  This gives 
\begin{align}
&G^{\delta}_1(a,\ta)=\frac{1}{\ta W(\ta)}\bigg(\frac{d D_{-}(\ta)}{d\ta}D_{+}(a)-\frac{d D_{+}(\ta)}{d\ta}D_{-}(a)\bigg)\Theta_{\rm H}(a-\ta)  \label{gdelta} \ , \\
&G^{\delta}_2(a,\ta)=\frac{f_{+}(\ta)/\ta^2}{W(\ta)}\bigg(D_{+}(\ta)D_{-}(a)-D_{-}(\ta)D_{+}(a)\bigg)\Theta_{\rm H}(a-\ta) \ , \\
&G^{\Theta}_1(a,\ta)=\frac{a/\ta}{f_{+}(a)W(\ta)}\bigg(\frac{d D_{-}(\ta)}{d\ta}\frac{d D_{+}(a)}{d a}-\frac{d D_{+}(\ta)}{d\ta}\frac{d D_{-}(a)}{d a}\bigg)\Theta_{\rm H}(a-\ta) \ ,\\
&G^{\Theta}_2(a,\ta)=\frac{f_{+}(\ta)a/\ta^2}{f_{+}(a)W(\ta)}\bigg(D_{+}(\ta)\frac{d D_{-}(a)}{d a}-D_{-}(\ta)\frac{d D_{+}(a)}{d a}\bigg)\Theta_{\rm H}(a-\ta) \ ,   \label{gtheta}
\end{align}
where $W(\ta)$ is the Wronskian of $D_+$ and $D_-$ 
\be
W(\ta)=\frac{dD_{-}(\ta)}{d\ta}D_{+}(\ta)-\frac{d D_{+}(\ta)}{d\ta}D_{-}(\ta) \ .
\ee
In addition, the counterterm is given by (see \cite{Lewandowski:2016yce} for details)
\be
\delta^{(ct)}_{\kvec} ( a ) = - ( 2 \pi ) \bar c_A^2 ( a ) \frac{k^2}{\knl^2} \frac{D_+(a)}{D_+(a_i)} \delta^{\rm in}_{\kvec}              \ . 
\ee

The expansion of the power spectrum is defined by  
\be
P ( a , k ) = P_{11} ( a , k ) + P_{22} ( a , k ) + P_{13} ( a , k ) + P_{13}^{ct} ( a , k )  + \cdots
\ee
where the various contributions are given by 
\begin{align}
 \langle\delta^{(1)}_{\vk}( a ) \delta^{(1)}_{\vk'}( a ) \rangle ' & = P_{11} ( a , k) \ , \\
 \langle\delta^{(2)}_{\vk}( a ) \delta^{(2)}_{\vk'}( a ) \rangle ' & =  P_{22} ( a , k) \ , \\
2   \langle\delta^{(1)}_{\vk}( a ) \delta^{(3)}_{\vk'}( a ) \rangle '  & =  P_{13} ( a , k) \ ,  \\
2   \langle\delta^{(1)}_{\vk}( a ) \delta^{(ct)}_{\vk'}( a ) \rangle '  & =  P_{13}^{ct} ( a , k) \ , 
 \end{align}
and $\langle \cdots \rangle ' $ means that we have removed a factor of $(2\pi)^3\delta_D(\vk+\vk')$ from the expectation value.  In particular, on the initial conditions, this means that $\langle \delta^{\rm in}_{\kvec} \delta^{\rm in}_{\kvec ' } \rangle' = P^{\rm in}_{\kvec}$.  This leads to the following expressions for the power spectrum contributions
\begin{align}
&P_{11}(a, k)=\frac{ D^2_{+}(a)  }{  D^2_+ ( a_i ) }P^{\rm in}_{\vk},\\  \label{P22}
&P_{22}(a , k)=2\int \frac{ d^3q}{(2 \pi)^3 } \bigg(\alpha_s(\vq,\vk-\vq)\mG^{\delta}_{1}(a)+\beta(\vq,\vk-\vq)\mG^{\delta}_{2}(a)\bigg)^2P^{\rm in}_{\vk-\vq}  \, P^{\rm in}_{\vq},\\\label{P13}
&P_{13}(a , k)= 4 \frac{ D_{+}(a)}{D_+ ( a_i)} P_{\vk}^{\rm in}\int \frac{d^3q}{ (2 \pi)^3} \bigg(\alpha^{\sigma}(\vk,\vk+\vq, \vk )\mU^{\delta}_{\sigma}(a)+\beta^{\sigma}(\vk,\vk+\vq,\vk)\mV^{\delta}_{\sigma2}(a) \nonumber \\ 
& \hspace{3.5in} +\gamma^{\sigma}(\vk,\vk+\vq,\vk)\mV^{\delta}_{\sigma1}(a)\bigg)P_{\vq}^{\rm in}  \ ,  \\
& P_{13}^{ct} ( a , k ) = - 2 \, ( 2 \pi ) \, \bar c_A^2 ( a ) \frac{k^2}{\knl^2} \left( \frac{ D_+ ( a ) }{D_+ ( a_i ) } \right)^2 P^{\rm in }_{\kvec} \ . 
\end{align}
Because the counterterm $P_{13}^{ct}$ is trivially IR safe, we will not consider it in this paper.  The momentum dependent functions in \eqn{P22} and \eqn{P13} are given as 
\begin{align}  \label{momentum-fun-3} 
\mathcal{\alpha}^1(\vk_{1},\vk_2,\vk_3)&\equiv\alpha(\vk_{1}-\vk_{2},\vk_{2})\alpha_s(\vk_{3},\vk_{2}-\vk_3) \ ,  \\ \mathcal{\alpha}^2(\vk_{1},\vk_2,\vk_3)&\equiv\alpha(\vk_{1}-\vk_{2},\vk_{2})\beta(\vk_{3},\vk_{2}-\vk_3) \ , \\
\mathcal{\beta}^1(\vk_{1},\vk_2,\vk_3)&\equiv2\beta(\vk_{1}-\vk_{2},\vk_{2})\alpha_s(\vk_{3},\vk_{2}-\vk_3) \ , \\ \mathcal{\beta}^2(\vk_{1},\vk_2,\vk_3)&\equiv2\beta(\vk_{1}-\vk_{2},\vk_{2})\beta(\vk_{3},\vk_{2}-\vk_3) \ , \\
\mathcal{\gamma}^1(\vk_{1},\vk_2,\vk_3)&\equiv\alpha(\vk_{2},\vk_{1}-\vk_{2})\alpha_s(\vk_{3},\vk_{2}-\vk_{3}) \ ,  \\ \mathcal{\gamma}^2(\vk_{1},\vk_2,\vk_3)&\equiv\alpha(\vk_{2},\vk_{1}-\vk_{2})\beta(\vk_{3},\vk_{2}-\vk_{3}) \label{momentum-fun-4}      \ , 
\end{align}
where $\alpha$, $\alpha_s$, and $\beta$ are the standard dark-energy interaction vertices given in \eqn{alphadef2} and \eqn{betadef2}.  The time dependent factors in \eqn{P22} and \eqn{P13} are given by 
\begin{align}    \label{2nd-solution} 
\mG^{\delta}_{\sigma}(a)&=\int^{1}_0 \frac{f_{+}(\ta)   D_{+}^2(\ta)  }{C(\ta) D_+^2 ( a_i) }G^{\delta}_{\sigma}(a,\ta)d\ta \ , \\
\mG^{\Theta}_{\sigma}(a)&=\int^{ 1}_0 \frac{f_{+}(\ta)  D_{+}^2(\ta)   }{C(\ta)  D_+^2 ( a_i ) }G^{\Theta}_{\sigma}(a,\ta)d\ta \ ,      \label{2nd-solution2} 
\end{align}
and
\begin{align} \label{timedepcoeff1}
\mU^{\delta}_{\sigma}(a)=\int^{{1}}_0 \frac{f_{+}(\ta)D_{+}(\ta)}{C(\ta) D_+ ( a_i) }\mG^{\delta}_{\sigma}(\ta)G^{\delta}_{1}(a,\ta)d\ta \ ,\\
\mU^{\Theta}_{\sigma}(a)=\int^{{1}}_0 \frac{f_{+}(\ta)D_{+}(\ta)}{C(\ta) D_+(a_i) }\mG^{\delta}_{\sigma}(\ta)G^{\Theta}_{1}(a,\ta)d\ta \ ,\\
\mV^{\delta}_{\sigma\tilde\sigma}(a)=\int^{{ 1}}_0 \frac{f_{+}(\ta)D_{+}(\ta)}{C(\ta)  D_+(a_i)}\mG^{\Theta}_{\sigma}(\ta)G^{\delta}_{\tilde\sigma}(a,\ta)d\ta \ ,\\
\mV^{\Theta}_{\sigma\tilde\sigma}(a)=\int^{{1}}_0 \frac{f_{+}(\ta)D_{+}(\ta)}{C(\ta)    D_+(a_i) }\mG^{\Theta}_{\sigma}(\ta)G^{\Theta}_{\tilde\sigma}(a,\ta)d\ta \ .  \label{timedepcoeff2}
\end{align}

\section{Fitting details} \label{fittingdetails}

In this appendix, we present the results of the fitting procedure used to determine the values of the counterterm parameters used in the precision comparison in Section~\ref{preccompsec}.  Although we refer the reader to~\cite{Foreman:2015lca} for details, we note that the curves from the determination of the parameters are slightly better when using the exact time dependence $P^{\text{IR\&UV-safe}}_{ 1-\text{loop}}$ that we presented in this paper, as they have smaller oscillations. 

\begin{figure}[htb!] 
\begin{center}
\includegraphics[width=14cm]{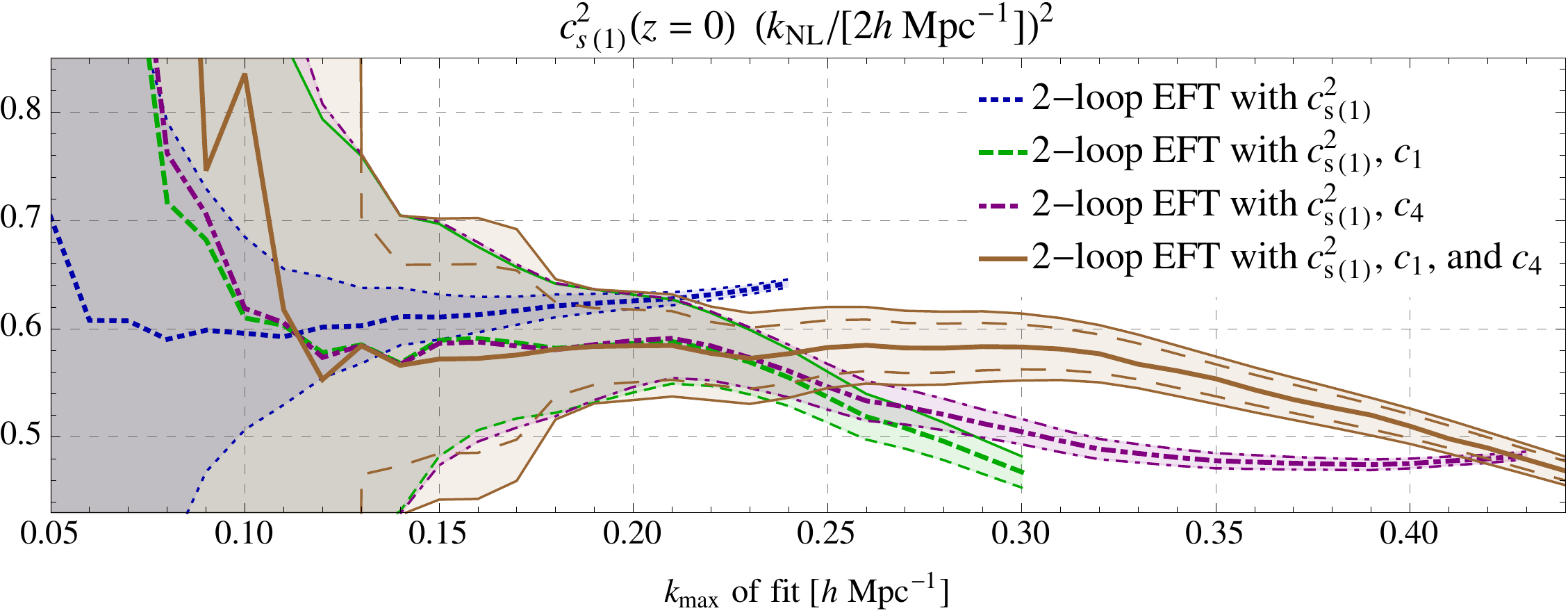} \\
\vspace{.2cm}
\includegraphics[width=14cm]{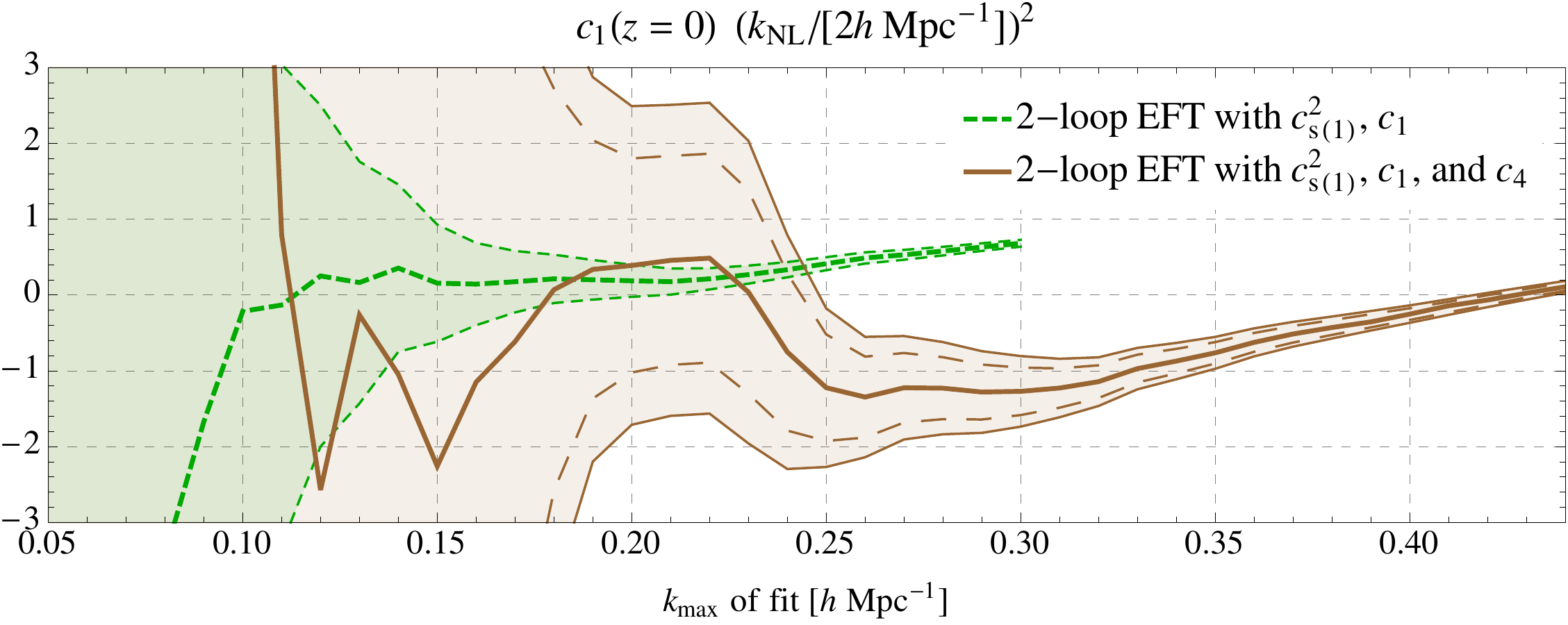} \\
\vspace{.2cm}
\includegraphics[width=14cm]{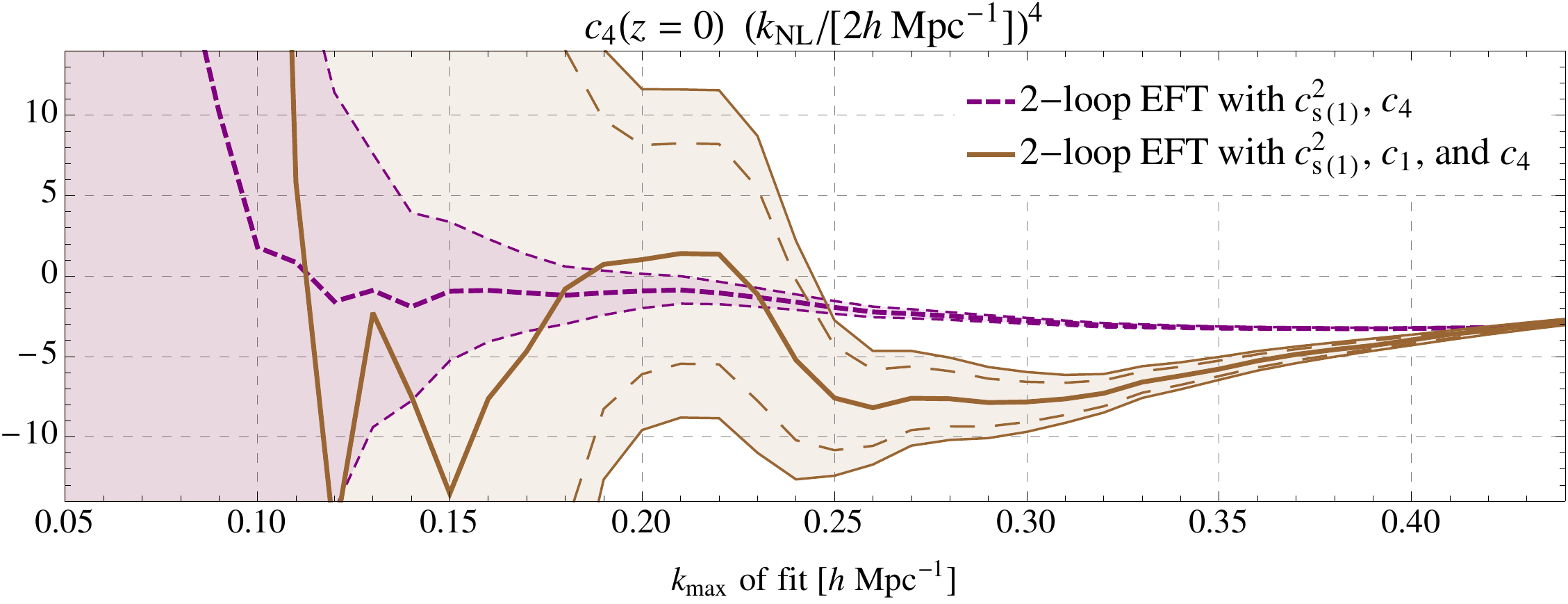} 
\caption{In this figure, we show the results of the fitting procedure used to determine the values of the various counterterm parameters.  These plots show the determined value of the counterterms $c_{s(1)}^2$, $c_1$, and $c_4$ as a function of the maximum $k$ used in the fit (called $k_{\rm max}$).  The shaded region is the $2 \sigma$ error region, and any long dashed lines represent a $1\sigma$ error region. }             \label{fit}
\end{center}
\end{figure}

{\footnotesize
\bibliography{references2}}

\providecommand{\href}[2]{#2}\begingroup\raggedright\begin{thebibliography}{10}

\bibitem{Baumann:2010tm}
D.~Baumann, A.~Nicolis, L.~Senatore, and M.~Zaldarriaga, {\it {Cosmological
  Non-Linearities as an Effective Fluid}},  {\em JCAP} {\bf 1207} (2012) 051,
  [\href{http://arxiv.org/abs/1004.2488}{{\tt arXiv:1004.2488}}].

\bibitem{Carrasco:2012cv}
J.~J.~M. Carrasco, M.~P. Hertzberg, and L.~Senatore, {\it {The Effective Field
  Theory of Cosmological Large Scale Structures}},  {\em JHEP} {\bf 09} (2012)
  082, [\href{http://arxiv.org/abs/1206.2926}{{\tt arXiv:1206.2926}}].

\bibitem{Porto:2013qua}
R.~A. Porto, L.~Senatore, and M.~Zaldarriaga, {\it {The Lagrangian-space
  Effective Field Theory of Large Scale Structures}},  {\em JCAP} {\bf 1405}
  (2014) 022, [\href{http://arxiv.org/abs/1311.2168}{{\tt arXiv:1311.2168}}].

\bibitem{Senatore:2014via}
L.~Senatore and M.~Zaldarriaga, {\it {The IR-resummed Effective Field Theory of
  Large Scale Structures}},  {\em JCAP} {\bf 1502} (2015) 013,
  [\href{http://arxiv.org/abs/1404.5954}{{\tt arXiv:1404.5954}}].

\bibitem{Carrasco:2013sva}
J.~J.~M. Carrasco, S.~Foreman, D.~Green, and L.~Senatore, {\it {The 2-loop
  matter power spectrum and the IR-safe integrand}},  {\em JCAP} {\bf 1407}
  (2014) 056, [\href{http://arxiv.org/abs/1304.4946}{{\tt arXiv:1304.4946}}].

\bibitem{Carrasco:2013mua}
J.~J.~M. Carrasco, S.~Foreman, D.~Green, and L.~Senatore, {\it {The Effective
  Field Theory of Large Scale Structures at Two Loops}},  {\em JCAP} {\bf 1407}
  (2014) 057, [\href{http://arxiv.org/abs/1310.0464}{{\tt arXiv:1310.0464}}].

\bibitem{Pajer:2013jj}
E.~Pajer and M.~Zaldarriaga, {\it {On the Renormalization of the Effective
  Field Theory of Large Scale Structures}},  {\em JCAP} {\bf 1308} (2013) 037,
  [\href{http://arxiv.org/abs/1301.7182}{{\tt arXiv:1301.7182}}].

\bibitem{Carroll:2013oxa}
S.~M. Carroll, S.~Leichenauer, and J.~Pollack, {\it {Consistent effective
  theory of long-wavelength cosmological perturbations}},  {\em Phys. Rev.}
  {\bf D90} (2014), no.~2 023518, [\href{http://arxiv.org/abs/1310.2920}{{\tt
  arXiv:1310.2920}}].

\bibitem{Mercolli:2013bsa}
L.~Mercolli and E.~Pajer, {\it {On the velocity in the Effective Field Theory
  of Large Scale Structures}},  {\em JCAP} {\bf 1403} (2014) 006,
  [\href{http://arxiv.org/abs/1307.3220}{{\tt arXiv:1307.3220}}].

\bibitem{Angulo:2014tfa}
R.~E. Angulo, S.~Foreman, M.~Schmittfull, and L.~Senatore, {\it {The One-Loop
  Matter Bispectrum in the Effective Field Theory of Large Scale Structures}},
  {\em JCAP} {\bf 1510} (2015) 039, [\href{http://arxiv.org/abs/1406.4143}{{\tt
  arXiv:1406.4143}}].

\bibitem{Baldauf:2014qfa}
T.~Baldauf, L.~Mercolli, M.~Mirbabayi, and E.~Pajer, {\it {The Bispectrum in
  the Effective Field Theory of Large Scale Structure}},  {\em JCAP} {\bf 1505}
  (2015), no.~05 007, [\href{http://arxiv.org/abs/1406.4135}{{\tt
  arXiv:1406.4135}}].

\bibitem{Senatore:2014eva}
L.~Senatore, {\it {Bias in the Effective Field Theory of Large Scale
  Structures}},  {\em JCAP} {\bf 1511} (2015), no.~11 007,
  [\href{http://arxiv.org/abs/1406.7843}{{\tt arXiv:1406.7843}}].

\bibitem{Senatore:2014vja}
L.~Senatore and M.~Zaldarriaga, {\it {Redshift Space Distortions in the
  Effective Field Theory of Large Scale Structures}},
  \href{http://arxiv.org/abs/1409.1225}{{\tt arXiv:1409.1225}}.

\bibitem{Lewandowski:2014rca}
M.~Lewandowski, A.~Perko, and L.~Senatore, {\it {Analytic Prediction of
  Baryonic Effects from the EFT of Large Scale Structures}},  {\em JCAP} {\bf
  1505} (2015) 019, [\href{http://arxiv.org/abs/1412.5049}{{\tt
  arXiv:1412.5049}}].

\bibitem{Mirbabayi:2014zca}
M.~Mirbabayi, F.~Schmidt, and M.~Zaldarriaga, {\it {Biased Tracers and Time
  Evolution}},  {\em JCAP} {\bf 1507} (2015), no.~07 030,
  [\href{http://arxiv.org/abs/1412.5169}{{\tt arXiv:1412.5169}}].

\bibitem{Foreman:2015uva}
S.~Foreman and L.~Senatore, {\it {The EFT of Large Scale Structures at All
  Redshifts: Analytical Predictions for Lensing}},  {\em JCAP} {\bf 1604}
  (2016) 033, [\href{http://arxiv.org/abs/1503.01775}{{\tt arXiv:1503.01775}}].

\bibitem{Angulo:2015eqa}
R.~Angulo, M.~Fasiello, L.~Senatore, and Z.~Vlah, {\it {On the Statistics of
  Biased Tracers in the Effective Field Theory of Large Scale Structures}},
  {\em JCAP} {\bf 1509} (2015) 029,
  [\href{http://arxiv.org/abs/1503.08826}{{\tt arXiv:1503.08826}}].

\bibitem{McQuinn:2015tva}
M.~McQuinn and M.~White, {\it {Cosmological perturbation theory in 1+1
  dimensions}},  {\em JCAP} {\bf 1601} (2016), no.~01 043,
  [\href{http://arxiv.org/abs/1502.07389}{{\tt arXiv:1502.07389}}].

\bibitem{Assassi:2015jqa}
V.~Assassi, D.~Baumann, E.~Pajer, Y.~Welling, and D.~van~der Woude, {\it
  {Effective theory of large-scale structure with primordial non-Gaussianity}},
   {\em JCAP} {\bf 1511} (2015) 024,
  [\href{http://arxiv.org/abs/1505.06668}{{\tt arXiv:1505.06668}}].

\bibitem{Baldauf:2015tla}
T.~Baldauf, E.~Schaan, and M.~Zaldarriaga, {\it {On the reach of perturbative
  descriptions for dark matter displacement fields}},  {\em JCAP} {\bf 1603}
  (2016), no.~03 017, [\href{http://arxiv.org/abs/1505.07098}{{\tt
  arXiv:1505.07098}}].

\bibitem{Baldauf:2015xfa}
T.~Baldauf, M.~Mirbabayi, M.~Simonovi\'{c}, and M.~Zaldarriaga, {\it
  {Equivalence Principle and the Baryon Acoustic Peak}},  {\em Phys. Rev.} {\bf
  D92} (2015), no.~4 043514, [\href{http://arxiv.org/abs/1504.04366}{{\tt
  arXiv:1504.04366}}].

\bibitem{Foreman:2015lca}
S.~{Foreman}, H.~{Perrier}, and L.~{Senatore}, {\it {Precision Comparison of
  the Power Spectrum in the EFTofLSS with Simulations}},  {\em JCAP} {\bf 1605}
  (2016) 027, [\href{http://arxiv.org/abs/1507.05326}{{\tt arXiv:1507.05326}}].

\bibitem{Baldauf:2015aha}
T.~Baldauf, L.~Mercolli, and M.~Zaldarriaga, {\it {Effective field theory of
  large scale structure at two loops: The apparent scale dependence of the
  speed of sound}},  {\em Phys. Rev.} {\bf D92} (2015), no.~12 123007,
  [\href{http://arxiv.org/abs/1507.02256}{{\tt arXiv:1507.02256}}].

\bibitem{Baldauf:2015zga}
T.~Baldauf, E.~Schaan, and M.~Zaldarriaga, {\it {On the reach of perturbative
  methods for dark matter density fields}},  {\em JCAP} {\bf 1603} (2016),
  no.~03 007, [\href{http://arxiv.org/abs/1507.02255}{{\tt arXiv:1507.02255}}].

\bibitem{Bertolini:2015fya}
D.~Bertolini, K.~Schutz, M.~P. Solon, J.~R. Walsh, and K.~M. Zurek, {\it
  {Non-Gaussian Covariance of the Matter Power Spectrum in the Effective Field
  Theory of Large Scale Structure}},
  \href{http://arxiv.org/abs/1512.07630}{{\tt arXiv:1512.07630}}.

\bibitem{Bertolini:2016bmt}
D.~Bertolini, K.~Schutz, M.~P. Solon, and K.~M. Zurek, {\it {The Trispectrum in
  the Effective Field Theory of Large Scale Structure}},
  \href{http://arxiv.org/abs/1604.01770}{{\tt arXiv:1604.01770}}.

\bibitem{Assassi:2015fma}
V.~Assassi, D.~Baumann, and F.~Schmidt, {\it {Galaxy Bias and Primordial
  Non-Gaussianity}},  {\em JCAP} {\bf 1512} (2015), no.~12 043,
  [\href{http://arxiv.org/abs/1510.03723}{{\tt arXiv:1510.03723}}].

\bibitem{Lewandowski:2015ziq}
M.~Lewandowski, L.~Senatore, F.~Prada, C.~Zhao, and C.-H. Chuang, {\it {On the
  EFT of Large Scale Structures in Redshift Space}},
  \href{http://arxiv.org/abs/1512.06831}{{\tt arXiv:1512.06831}}.

\bibitem{Cataneo:2016suz}
M.~Cataneo, S.~Foreman, and L.~Senatore, {\it {Efficient exploration of
  cosmology dependence in the EFT of LSS}},
  \href{http://arxiv.org/abs/1606.03633}{{\tt arXiv:1606.03633}}.

\bibitem{Bertolini:2016hxg}
D.~Bertolini and M.~P. Solon, {\it {Principal Shapes and Squeezed Limits in the
  Effective Field Theory of Large Scale Structure}},
  \href{http://arxiv.org/abs/1608.01310}{{\tt arXiv:1608.01310}}.

\bibitem{Fujita:2016dne}
T.~Fujita, V.~Mauerhofer, L.~Senatore, Z.~Vlah, and R.~Angulo, {\it {Very
  Massive Tracers and Higher Derivative Biases}},
  \href{http://arxiv.org/abs/1609.00717}{{\tt arXiv:1609.00717}}.

\bibitem{Perko:2016puo}
A.~Perko, L.~Senatore, E.~Jennings, and R.~H. Wechsler, {\it {Biased Tracers in
  Redshift Space in the EFT of Large-Scale Structure}},
  \href{http://arxiv.org/abs/1610.09321}{{\tt arXiv:1610.09321}}.

\bibitem{Lewandowski:2016yce}
M.~Lewandowski, A.~Maleknejad, and L.~Senatore, {\it {An effective description
  of dark matter and dark energy in the mildly non-linear regime}},
  \href{http://arxiv.org/abs/1611.07966}{{\tt arXiv:1611.07966}}.

\bibitem{Jain:1995kx}
B.~Jain and E.~Bertschinger, {\it {Selfsimilar evolution of cosmological
  density fluctuations}},  {\em Astrophys. J.} {\bf 456} (1996) 43,
  [\href{http://arxiv.org/abs/astro-ph/9503025}{{\tt astro-ph/9503025}}].

\bibitem{Scoccimarro:1995if}
R.~Scoccimarro and J.~Frieman, {\it {Loop corrections in nonlinear cosmological
  perturbation theory}},  {\em Astrophys. J. Suppl.} {\bf 105} (1996) 37,
  [\href{http://arxiv.org/abs/astro-ph/9509047}{{\tt astro-ph/9509047}}].

\bibitem{Peloso:2013zw}
M.~Peloso and M.~Pietroni, {\it {Galilean invariance and the consistency
  relation for the nonlinear squeezed bispectrum of large scale structure}},
  {\em JCAP} {\bf 1305} (2013) 031, [\href{http://arxiv.org/abs/1302.0223}{{\tt
  arXiv:1302.0223}}].

\bibitem{Kehagias:2013yd}
A.~Kehagias and A.~Riotto, {\it {Symmetries and Consistency Relations in the
  Large Scale Structure of the Universe}},  {\em Nucl. Phys.} {\bf B873} (2013)
  514--529, [\href{http://arxiv.org/abs/1302.0130}{{\tt arXiv:1302.0130}}].

\bibitem{Bernardeau:2001qr}
F.~Bernardeau, S.~Colombi, E.~Gaztanaga, and R.~Scoccimarro, {\it {Large scale
  structure of the universe and cosmological perturbation theory}},  {\em Phys.
  Rept.} {\bf 367} (2002) 1--248,
  [\href{http://arxiv.org/abs/astro-ph/0112551}{{\tt astro-ph/0112551}}].

\bibitem{Bernardeau:2011vy}
F.~Bernardeau, N.~Van~de Rijt, and F.~Vernizzi, {\it {Resummed propagators in
  multi-component cosmic fluids with the eikonal approximation}},  {\em Phys.
  Rev.} {\bf D85} (2012) 063509, [\href{http://arxiv.org/abs/1109.3400}{{\tt
  arXiv:1109.3400}}].

\bibitem{Bernardeau:2012aq}
F.~Bernardeau, N.~Van~de Rijt, and F.~Vernizzi, {\it {Power spectra in the
  eikonal approximation with adiabatic and nonadiabatic modes}},  {\em Phys.
  Rev.} {\bf D87} (2013), no.~4 043530,
  [\href{http://arxiv.org/abs/1209.3662}{{\tt arXiv:1209.3662}}].

\bibitem{Creminelli:2013mca}
P.~Creminelli, J.~Nore\~{n}a, M.~Simonovi\'{c}, and F.~Vernizzi, {\it
  {Single-Field Consistency Relations of Large Scale Structure}},  {\em JCAP}
  {\bf 1312} (2013) 025, [\href{http://arxiv.org/abs/1309.3557}{{\tt
  arXiv:1309.3557}}].

\bibitem{Weinberg:2003sw}
S.~Weinberg, {\it {Adiabatic modes in cosmology}},  {\em Phys. Rev.} {\bf D67}
  (2003) 123504, [\href{http://arxiv.org/abs/astro-ph/0302326}{{\tt
  astro-ph/0302326}}].

\bibitem{Tseliakhovich:2010bj}
D.~Tseliakhovich and C.~Hirata, {\it {Relative velocity of dark matter and
  baryonic fluids and the formation of the first structures}},  {\em Phys.
  Rev.} {\bf D82} (2010) 083520, [\href{http://arxiv.org/abs/1005.2416}{{\tt
  arXiv:1005.2416}}].

\bibitem{Takahashi:2008yk}
R.~Takahashi, {\it {Third Order Density Perturbation and One-loop Power
  Spectrum in a Dark Energy Dominated Universe}},  {\em Prog. Theor. Phys.}
  {\bf 120} (2008) 549--559, [\href{http://arxiv.org/abs/0806.1437}{{\tt
  arXiv:0806.1437}}].

\bibitem{martel1991}
H.~Martel and W.~Freudling, {\it {Second-order perturbation theory in Omega is
  not equal to Friedmann models}},  {\em Astrophysics J.} {\bf 371} (1991) 1.

\bibitem{Bernardeau:1993qu}
F.~Bernardeau, {\it {Skewness and Kurtosis in large scale cosmic fields}},
  {\em Astrophys. J.} {\bf 433} (1994) 1,
  [\href{http://arxiv.org/abs/astro-ph/9312026}{{\tt astro-ph/9312026}}].

\bibitem{Scoccimarro:1997st}
R.~Scoccimarro, S.~Colombi, J.~N. Fry, J.~A. Frieman, E.~Hivon, and A.~Melott,
  {\it {Nonlinear evolution of the bispectrum of cosmological perturbations}},
  {\em Astrophys. J.} {\bf 496} (1998) 586,
  [\href{http://arxiv.org/abs/astro-ph/9704075}{{\tt astro-ph/9704075}}].

\bibitem{Skillman:2014qca}
S.~W. Skillman, M.~S. Warren, M.~J. Turk, R.~H. Wechsler, D.~E. Holz, and P.~M.
  Sutter, {\it {Dark Sky Simulations: Early Data Release}},
  \href{http://arxiv.org/abs/1407.2600}{{\tt arXiv:1407.2600}}.

\bibitem{Creminelli:2008wc}
P.~Creminelli, G.~D'Amico, J.~Nore\~{n}a, and F.~Vernizzi, {\it {The Effective
  Theory of Quintessence: the w$<$-1 Side Unveiled}},  {\em JCAP} {\bf 0902}
  (2009) 018, [\href{http://arxiv.org/abs/0811.0827}{{\tt arXiv:0811.0827}}].

\bibitem{Creminelli:2009mu}
P.~Creminelli, G.~D'Amico, J.~Nore\~{n}a, L.~Senatore, and F.~Vernizzi, {\it
  {Spherical collapse in quintessence models with zero speed of sound}},  {\em
  JCAP} {\bf 1003} (2010) 027, [\href{http://arxiv.org/abs/0911.2701}{{\tt
  arXiv:0911.2701}}].

\bibitem{Sefusatti:2011cm}
E.~Sefusatti and F.~Vernizzi, {\it {Cosmological structure formation with
  clustering quintessence}},  {\em JCAP} {\bf 1103} (2011) 047,
  [\href{http://arxiv.org/abs/1101.1026}{{\tt arXiv:1101.1026}}].

\bibitem{Blas:2013bpa}
D.~Blas, M.~Garny, and T.~Konstandin, {\it {On the non-linear scale of
  cosmological perturbation theory}},  {\em JCAP} {\bf 1309} (2013) 024,
  [\href{http://arxiv.org/abs/1304.1546}{{\tt arXiv:1304.1546}}].

\end{thebibliography}\endgroup

\end{document}